\documentclass[twocolumn,amsmath,amssymb,pra]{revtex4}

\usepackage{graphics}
\usepackage{epsfig}
\usepackage{amsmath}
\usepackage{color}
\usepackage{dcolumn}
\usepackage{bm}

\newcommand{\ds }{\displaystyle}

\newcommand{\eci }{\mathcal{E}}
\newcommand{\RbK} {$^{87}$Rb-$^{40}$K }
\newcommand{\LiLi}{$^{6}$Li-$^{7}$Li }
\newcommand{\LiCs}{$^{6}$Li-$^{133}$Cs }
\newcommand{\ket}[1]{| #1 \rangle}

\def\oper#1{\hat{\rm #1}}

\begin{document}

  \bibliographystyle{apsrev}

  \title{Exact theoretical description of two ultracold atoms in a 
         single site of a 3D optical lattice using realistic 
         interatomic interaction potentials}

       \author{Sergey Grishkevich and Alejandro Saenz}

       \affiliation{AG Moderne Optik, Institut f\"ur Physik,
         Humboldt-Universit\"at zu Berlin, Hausvogteiplatz 5-7,
         10117 Berlin, Germany}

       \date{\today}

  \begin{abstract}
     A theoretical approach was developed for an exact numerical description 
     of a pair of ultracold atoms interacting via a central potential that are  
     trapped in a three-dimensional optical lattice. The coupling of 
     center-of-mass and relative-motion coordinates is explicitly considered 
     using a configuration-interaction (exact-diagonalization) technique. 
     Deviations from the harmonic approximation are discussed for several 
     heteronuclear alkali-metal atom pairs trapped in a single site of an 
     optical lattice. The consequences are discussed for the analysis of a 
     recent experiment [C.~Ospelkaus {\it et al}, Phys.\,Rev.\,Lett.~{\bf 97},
     120402 (2006)] 
     in which radio-frequency association was used to create diatomic molecules 
     from a fermionic and a bosonic atom and to measure their binding energies
     close to a magnetic Feshbach resonance.  
  \end{abstract}

    \maketitle

\section{Introduction}
\label{sec:intro}
 The physics of ultracold atoms has attracted a lot of interest since 
 the experimental observation of Bose-Einstein condensation in dilute 
 alkali-metal atom gases~\cite{cold:ande95,cold:davi95}. Besides the exciting 
 physics at ultracold energies by itself, a further important progress 
 was the loading of the ultracold gas into an optical lattice formed with
 the aid of standing light waves~\cite{cold:jaks98,cold:grei02,cold:koeh05}. 
 The optical lattice resembles in some sense the periodicity of a crystal
 potential~\cite{cold:meac98,cold:bloc05,cold:lewe07}. In contrast to real
 solids the lattice parameters are, however, easily tunable by a variation
 of the laser intensity (trap depth) or wavelength (lattice geometry).
 Moreover, different atoms possess different interaction potentials
 that can be either attractive or repulsive. While different
 kinds of chemical elements, their isotopes, or atoms in different electronic 
 or spin states cover already quite some range of interaction strengths, a further
 tunability of the atom-atom interactions in optical lattices can be achieved 
 using magnetic Feshbach resonances~\cite{cold:loft02,cold:rega03}. Close to the
 resonance value of the magnetic field the interaction varies in a wide range
 of attractive and repulsive values.

 Ultracold atoms deposited in light crystals are ideal systems for a realization
 of the Hubbard model~\cite{cold:jaks98}. This model takes into account a single
 band of a static lattice potential and assumes the interactions to be purely
 local~\cite{cold:hubb63} (low tunneling limit). In this case the optical
 lattice is considered as an array with a very small filling rate; optimally 
 with one or two atoms per lattice site. The experimental study of a bosonic Mott
 insulator~\cite{cold:grei02} and a fermionic band
 insulator~\cite{cold:koeh05} provided such a system. The fact that the analysis
 of a single site is sufficient for such systems simplifies their theoretical
 description drastically.

 Another interesting aspect is that ultracold atoms can bind together to form
 ultracold molecules. The optical lattice can shield the often fragile,
 very weakly bound molecules from destructive three-body collisions.
 The physics of ultracold atom pairs in optical lattices
 with controllable interactions is thus presently an intensively investigated
 research area. Recently the observation of confinement-induced molecules,
 repulsively interacting pairs, and real molecules for both
 homonuclear~\cite{cold:koeh05,cold:stoe06} and
 heteronuclear~\cite{cold:ospe06a} atomic species in optical lattices has been
 reported.

 In order to describe the behavior of atoms in an optical lattice the latter
 is usually considered as an array of harmonic traps. In such an approach some
 important features of the optical lattice can be lost. For example, the
 correct sinusoidal potential exhibits an energy band with a spread of
 transition energies while the harmonic potential possesses a discrete
 equidistant spectrum. Nevertheless, the experiment of St\"oferle et
 al.~\cite{cold:stoe06} showed good agreement with a simplified theoretical
 description based on the harmonic approximation. In their analysis,
 St\"oferle et al.\ compared the measured binding energies of
 confinement-induced molecules and real molecules to the ones predicted by a
 simplified theory in which two atoms are trapped in a harmonic potential and
 interact {\it via} a $\delta$-function pseudopotential. Within such a model
 an analytical solution exists in the case of two identical atoms (in the same
 quantum states) \cite{cold:busc98}. However, another experiment that adopted
 higher resolution spectroscopy and considered a heteronuclear system was
 interpreted to clearly indicate a break-down of the harmonic
 approximation~\cite{cold:ospe06a}.

 From the theoretical point of view, the description of two atoms in an  
 optical lattice beyond the harmonic approximation is very laborious. 
 The anharmonic part of the optical lattice potential leads to a coupling 
 of center-of-mass and relative motion and requires therefore to solve 
 the full six-dimensional problem. In fact, even within the harmonic 
 approximation different trapping potentials experienced by the two atoms 
 lead to a coupling of center-of-mass and relative 
 motion~\cite{cold:bold05,cold:gris07}. This situation occurs, e.\,g., for
 heteronuclear atom pairs or two atoms of the same kind but in different
 electronic states.

 In this work a numerical approach is developed that allows in principle to
 describe two atoms trapped in an optical lattice in an exact way, if the 
 interatomic interaction potential is central (isotropic) and can be given 
 in terms of a single potential curve. The fact that this latter curve stems 
 usually from a full molecular calculation and is only given in numerical form
 is explicitly considered. Extensions to non-central (like dipolar)
 interactions are rather straightforward and planned for the near future. 
 The anharmonic coupling is treated in a configuration-interaction (CI) like 
 fashion also known as exact diagonalization. It leads for sufficiently large
 expansion lengths to exact results. Although the approach allows to consider 
 multiple-well potentials and thus more than a single site of an optical 
 lattice~\cite{cold:schn09}, the present work focuses on results obtained for two atoms in 
 a single site. Motivated by the experiment reported in~\cite{cold:ospe06a} 
 the heteronuclear atom pair formed by fermionic $^{40}$K and bosonic 
 $^{87}$Rb was used as a generic system in the present study. In order to 
 investigate the influence of the atomic interaction strength, its value 
 was varied artificially by a controlled manipulation of the inner wall 
 of the corresponding potential curve. For the investigation of the 
 influence of the mass difference on the results, two other heteronuclear 
 pairs, $^6$Li-$^7$Li (almost equal masses) and $^6$Li-$^{133}$Cs 
 (very large mass difference) were considered. After a systematic
 investigation of the effects of anharmonicity and coupling of center-of-mass 
 and relative motion a comparison to the experimental data~\cite{cold:ospe06a}
 as well as to a subsequent theoretical analysis~\cite{cold:deur08} performed
 in parallel to the present work is given.  

 The paper is organized in the following way. In Sec.\,\ref{sec:system} 
 the theoretical approach is described. This includes the formulation of the 
 problem in Sec.\,\ref{subsec:hamilt} as well as a description of the trap 
 parameters (Sec.\,\ref{subsec:traparam}), the used interatomic interaction 
 potentials (Sec.\,\ref{subsec:inin}), and its systematic variation in 
 Sec.\,\ref{subsec:manofinin}. The section ends with the computational 
 details described in Sec.\,\ref{subsec:compdet}. The results presented 
 in Sec.\,\ref{sec:results} are first discussed in terms of energies (and 
 their differences) for the generic \RbK dimer in Sec.\,\ref{subsec:spectrum})
 and for other dimers in Sec.\,/\ref{subsec:othsys}. This is followed by 
 an analysis of the radial pair densities (Sec.\,\ref{subsec:raddens}) and 
 the wavefunctions in absolute coordinates (Sec.\,\ref{subsec:wfabsolute}). 
 This is followed by a comparison to the 
 experimental and recent alternative theoretical results in 
 Sec.\,\ref{subsec:compthexp}. 
 Finally, a conclusion and outlook is given in Sec.\,\ref{sec:outlook}. 
 All equations and quantities in this paper
 are given in atomic units unless otherwise specified.

\section{System}
\label{sec:system}
\subsection{Hamiltonian}
\label{subsec:hamilt}
 The Hamiltonian describing the interaction of two atoms with coordinate 
 vectors $\vec{r}_1$ and $\vec{r}_2$ trapped in a three-dimensional optical 
 lattice is given by 
 \begin{eqnarray}
   \label{eq:origham}
     \oper{H}(\vec{r}_1,\vec{r}_2)
           &=&
           \oper{T}_{1}(\vec{r}_1)
           + \oper{T}_{2}(\vec{r}_2)
           + \oper{U}(\vec{r}_1,\vec{r}_2)
           \nonumber \\
           &+&
           \oper{V}_{\rm trap,1} (\vec{r}_1)
           + \oper{V}_{\rm trap,2} (\vec{r}_2)
 \end{eqnarray}
 where $\oper{T}_j$ is the kinetic energy operator for particle
 $j$, $\oper{U}$ is the atom-atom interaction potential, and
 $\oper{V}_{\rm trap,j}$ is the trapping potential for particle $j$. 
 For optical lattices $\oper{V}_{\rm trap,j}$ is often (and also in the
 present work) given by
\begin{equation}
   \label{eq:sinform}
   \oper{V}_{\rm trap,j} =
           \sum\limits_{c={x,y,z}}   V_c^j \: \sin^2(k_c c_j)\, ,
 \end{equation}
 where $V_c^j$ is the potential depth which particle $j$ experiences along
 direction $c$, and $k_c=2\pi/\lambda_c$ is the wave vector and 
 $\lambda_c$ the wavelength of the laser creating the lattice potential 
 along the (Cartesian) coordinate $c$.   

 A direct solution of the Schr\"odinger equation with the
 Hamiltonian given in the form of Eq.\,(\ref{eq:origham}) is complicated, 
 since $\oper{U}$ depends in general on all six coordinates describing the 
 two-particle system, even if the atom-atom interaction is central, 
 i.\,e.\ $\oper{U}= \oper{U} (r)$ with $r=|\vec{r}_1 - \vec{r}_2|$. For
 realistic interatomic interaction potentials (that are usually even only
 known numerically), there is no separability and this leads to very demanding
 six-dimensional integrals. Therefore, it is more convenient to treat the
 two-particle problem in center-of-mass (COM) and relative (REL) motion
 coordinates. If spherical coordinates are adopted, a central interaction
 potential leads to a function of the radial coordinate only. 

 On the other hand, the formulation of the two-particle problem in COM and REL
 coordinates complicates the treatment of the trap potential, since its 
 separability in Cartesian coordinates is lost in the COM and REL coordinate 
 system. However, performing a Taylor expansion of the sinusoidal
 trapping potential~(\ref{eq:sinform}) around the origin simplifies the
 problem drastically, because the angular parts can be analytically solved  
 for in the case of a central interatomic interaction potential. For  
 two identical atoms in the same state the use of the harmonic 
 approximation for the trapping potential leads even to a problem 
 that is completely separable in COM and REL coordinates~\cite{cold:blum02}. 
 If the true atom-atom interaction is, furthermore, replaced by a
 $\delta$-function pseudopotential that reproduces only asymptotically the 
 two-body zero-energy s-wave scattering, the Schr\"odinger equation possesses an
 analytical solution for both isotropic or anisotropic harmonic
 traps~\cite{cold:busc98,cold:idzi05}. Noteworthy, even within the harmonic
 approximation the separability is lost, if the two atoms experience different
 trapping potentials. This is the case, if a heteronuclear system or two
 identical atoms in different electronic states are considered. 
  
 After performing the Taylor expansion of the sinusoidal trapping
 potential~(\ref{eq:sinform}) around the origin, the transformation of the
 Hamiltonian~(\ref{eq:origham}) into the COM and REL coordinate 
 systems leads to a Hamiltonian of the form
 \begin{equation}
   \label{eq:fullham}
   \oper{H}(\vec{R},\vec{r}\,) = \oper{h}_{\rm COM}(\vec{R}\,) 
                              + \oper{h}_{\rm REL}(\vec{r}\,)
                              + \oper{W}(\vec{R},\vec{r}\,) 
 \end{equation}
  with
    \begin{eqnarray}
      \ds &\,& \oper{h}_{\rm COM}(\vec{R}\,) = \oper{t}_{\rm kin}(\vec{R}\,)
                                + \oper{v}_{\rm OL}(\vec{R}\,),
                                \label{eq:hamcm}
           \\ 
          &\,& \oper{h}_{\rm REL}(\vec{r}\,) = \oper{T}_{\rm kin}(\vec{r}\,)
                                + \oper{V}_{\rm OL}(\vec{r}\,)
                                + \oper{V}_{\rm int}(\vec{r}\,)
                                \label{eq:hamrm}   \quad .
      \end{eqnarray}
 It is worth emphasizing that in the present formulation only the
 truly non-separable terms (represented by products of COM and REL
 coordinates) are left in the coupling term $\oper{W}$. All separable 
 terms of the optical lattice (OL) potential are included into the COM and
 REL Hamiltonians $\oper{h}_{\rm COM}$ and $\oper{h}_{\rm REL}$ respectively.

 In a first step, the eigenstates and -values of the COM and REL 
 Hamiltonians are obtained independently of each other by means of 
 a numerical solution of 
 the corresponding stationary Schr\"odinger equations,%
\begin{equation}
\label{eq:COMRELSchroed}
   \oper{h}_{\rm COM} \,|\psi_i\rangle = \varepsilon_i \,|\psi_i\rangle
   \quad {\rm and} \quad 
   \oper{h}_{\rm REL} \,|\phi_i\rangle = \epsilon_i \,|\phi_i\rangle \quad .    
\end{equation}
  The wavefunctions $\psi(\vec{R}\,)$ and $\phi(\vec{r}\,)$ are then used 
  to form the configuration state functions $\Phi_k(\vec{R},\vec{r}\,) =
  \psi_{i_k}(\vec{R}\,) \, \phi_{j_k}(\vec{r}\,)$. The stationary
  Schr\"odinger equation with the full Hamiltonian given in 
  Eq.\,(\ref{eq:fullham}),%
\begin{equation}
\label{eq:fullSchroed}
  \oper{H}\,|\Psi_i\rangle = \eci_i \,|\Psi_i\rangle \quad ,    
\end{equation}
 is then solved by expanding $\Psi$ as $\Psi(\vec{R},\vec{r}\,) =
 \sum_{k} \, \tilde{C}_k \, \Phi_k (\vec{R},\vec{r}\,)$. Insertion of 
 this expansion into Eq.\,(\ref{eq:fullSchroed}) leads (after the usual 
 manipulations) to a matrix eigenvalue problem that is solved numerically 
 and yields the energies $\eci_i$ and eigenvector coefficients 
 $\tilde{C}_k$.

\subsection{Trap parameters}
\label{subsec:traparam}
 Despite the already mentioned breakdown of the harmonic approximation 
 especially for heteronuclear systems it is still convenient to introduce 
 the mean harmonic-oscillator frequencies $\omega_{\rm ho}$ and 
 $\Omega_{\rm ho}$ of a single lattice site for the REL and COM motion
 respectively,
 \begin{eqnarray}
   \label{eq:omegaheteronuc}
   \ds \omega_{\rm{ho}} &=& k \; \sqrt{ 2
                \frac{V_1 {\mu_2}^2 + 
                V_2 {\mu_1}^2}
                     {\mu} } \\
   \label{eq:omegaheteronucCOM}
   \ds \Omega_{\rm ho} &=& k \; \sqrt{ 2
                \frac{V_1 + 
                V_2}
                     {M} } \quad .
 \end{eqnarray}
 In Eqs.~(\ref{eq:omegaheteronuc}) and (\ref{eq:omegaheteronucCOM}) $\mu$ and
 $M$ denote the reduced mass and total mass of the two particles
 respectively, $\mu_j$ is defined as $\mu_{1,2}=\mu/m_{2,1}$ where $m_j$ is
 the mass of atom $j$, and $\ds V_j = I_0\cdot\alpha_j$ is the optical lattice
 depth that is equal to the product of the laser intensity $I_0$ (for an
 isotropic geometry $I_0=I_x=I_y=I_z$) and the polarizabilities $\alpha_j$ of
 atom $j$. Finally, one has $k=k_x=k_y=k_z$ for an isotropic geometry of the
 lattice. This isotropy is in fact assumed in
 Eqs.~(\ref{eq:omegaheteronuc}) and (\ref{eq:omegaheteronucCOM}). For identical
 particles of mass $m$ Eq.~(\ref{eq:omegaheteronuc}) reduces to the well-known
 relation $\ds \omega_{\rm{ho}} = k \sqrt{2 V_0/m} $~\cite{cold:bold05}.

 Some parameters of the trap chosen in the present study were motivated by the 
 recent experiment reported in~\cite{cold:ospe06a}. Therein a
 three-dimensional optical lattice generated by lasers with wavelength
 $\lambda=\lambda_x=\lambda_y=\lambda_z$ of 1030\,nm was used for the trapping of ultracold bosonic $^{87}$Rb
 and fermionic $^{40}$K atoms. The two different lattice depths 
 $V_{\rm{Rb}}=40\,E_r^{\rm{Rb}}$ and $V_{\rm{Rb}}=27.5\,E_r^{\rm{Rb}}$ were
 considered where the individual recoil energy is defined, e.g., as
 $E_r^{\rm{Rb}}=k^2/(2m_{\rm{Rb}})$. Since the static dipole polarizabilities of
 rubidium and potassium are different, $\alpha_{\rm{Rb}} = 324$\,a.\,u.\ and  
 $\alpha_{\rm{K}} =\,301$\,a.\,u.~\cite{vdw:lim99}, the two atoms experience 
 different potentials: $V_{\rm{K}}=37.2\,E_r^{\rm{Rb}}$ and
 $V_{\rm{K}}=25.5\,E_r^{\rm{Rb}}$ for $40\,E_r^{\rm{Rb}}$ and
 $27.5\,E_r^{\rm{Rb}}$ respectively. The mean harmonic-oscillator
 frequencies~(\ref{eq:omegaheteronuc}) are 
 $\omega_{\rm{ho}}(40\,E_r^{\rm{Rb}}) = 2\pi\times
 35.7$~kHz and $\omega_{\rm{ho}}(27.5\,E_r^{\rm{Rb}}) = 2\pi\times 30$~kHz. 
 While most of the results of this work are obtained for these 
 frequencies, some other values are also considered in order to investigate 
 the influence of the trap frequency in more detail.  

\subsection{Interatomic interaction potential}
\label{subsec:inin}
 The interaction between rubidium and potassium atoms is modeled using
 the Born-Oppenheimer (BO) potential of the $a\, ^3\Sigma^+$ electronic state
 describing the interaction of two spin-polarized atoms. In general the
 atom-atom interaction potentials are only known numerically. For the
 short-range part $V_{\rm{SR}}$ of the potential in between $R\in [1.588 \,
 a_0, 18.2 \, a_0]$ the data of~\cite{vdw:rous00} are used ($a_0$ is the Bohr
 radius). The data points at $R=17.6 \,a_0$ and $R=16.99998 \,a_0$ have been
 omitted, because their inclusion results in a non-smooth potential curve. 
 The long range part $V_{\rm{LR}}$ of the $a\, ^3\Sigma^+$ electronic state 
 is constructed in a similar way as was done
 by Zemke et al.~\cite{vdw:zemk05}. Therefore, this long range part is 
 defined as $V_{\rm{LR}}(r)
 = D_e + \Delta V_{\rm{disp}}(r) + \Delta V_{\rm{ex}}(r)$ for $R \ge 18.6
 \, a_0$ where $\Delta V_{\rm{disp}}(r) = -C_6/r^6-C_8/r^8-C_{10}/r^{10}$
 and the dispersion coefficients $C_n$ are the values of Derevianko
 and co-workers~\cite{vdw:dere01,vdw:pors03} except $C_6=4292\pm 19$\,a.u.\
 which was taken from~\cite{vdw:ferl06}. The exchange interaction is given by
 $\Delta V_{\rm{ex}}(r) = -C r^{\alpha}e^{-\beta r}$ with $C=0.00231382$,
 $\alpha = 5.25603$, $\beta = 1.11892$ as given in~\cite{vdw:zemk05}. To merge
 the short- and the long-range parts the short-range part is raised up by 
 half of the value $\ds  \delta_{\rm{merge}} = V_{\rm{SR}}(18.2\, a_0) -
 V_{\rm{LR}}(18.6\, a_0)$. According to~\cite{vdw:zemk05} the $a ^3\Sigma^+$ 
 state supports 32 bound states and the interaction of the atoms via the $a 
 ^3\Sigma^+$ potential is strong and repulsive. The same amount of bound
 states and the same character of the interaction are observed in the present
 calculation using the potential curve constructed the way described above.

 In order to study the influence of different masses, polarizabilities,
 different interaction potentials, and also to check the generality of the 
 conclusions of this work other systems are also analyzed. 
 In particular, the heteronuclear $^6$Li-$^7$Li and $^6$Li-$^{133}$Cs pairs 
 interacting via their respective $a\, ^3\Sigma^+$ electronic state 
 are considered. The potential curves for $^6$Li-$^7$Li and $^6$Li-$^{133}$Cs 
 were constructed according to~\cite{cold:gris07} and~\cite{vdw:staa07}, 
 respectively.

\subsection{Manipulation of the interatomic interaction}
\label{subsec:manofinin}
\begin{figure}[ht]
 \centering
\includegraphics[width=7.5cm,height=6cm]{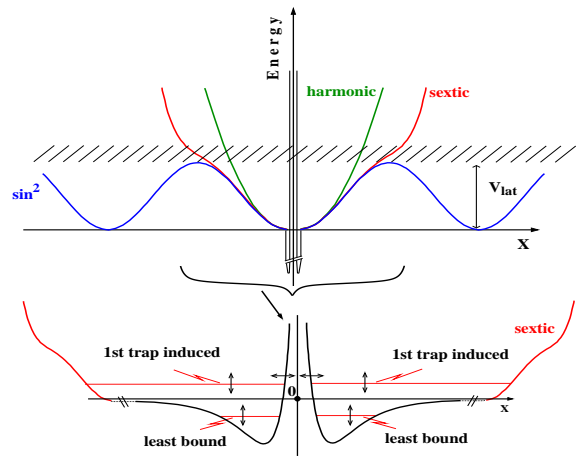}
 \caption{{\footnotesize
          (Color online) 
          Sketch (not to scale) of a cut through the 
          potential surfaces along the $x$ direction 
          ($y=z=0$) for a system of two {\it identical} atoms
          if one of them is positioned at the zero of $x$.
          The upper graph shows the $\sin^2$ potential together 
          with the harmonic (green) and sextic (red) approximations.
          While the harmonic and sextic potentials 
          support solely bound states, the energy spectrum of 
          the $\sin^2$ potential is partly discrete (for a 
          sufficiently deep value of $V_{\rm lat}$) and partly 
          continuous. The lower graph shows the range of small 
          $x$ values on an enlarged scale.
          A very tiny variation of the inner wall of the
          interaction potential leads to a relatively 
          large shift of the least bound and the first trap-induced
          states.
} }
\label{fig:sketch}
\end{figure}

 In the limit of zero collision energy the interaction between two atoms can
 be characterized by their s-wave scattering length $\ds a_{\rm sc}$. The sign
 of $\ds a_{\rm sc}$ determines the type of interaction (repulsive or
 attractive) and the absolute value determines the interaction
 strength. Experimentally it is difficult to accurately measure the scattering
 length. For example, there is no agreement about the value of the triplet
 scattering length for the \RbK system. According to the ongoing
 discussion~\cite{vdw:gold04,vdw:inou04,vdw:ferl06,cold:zacc06} the value
 $-185(4)\,a_0$ appears to be the most reliable one. A standard way to 
 match the calculated scattering length with the experimental value is a 
 smooth shift of the inner wall of the BO potential as is described 
 in~\cite{vdw:zemk05}. This 
 procedure can also be used for an effective variation of the scattering
 length since a systematic variation of the inner wall allows to shift the
 least bound (lb) state supported by the potential curve as is shown in
 Fig.~\ref{fig:sketch}. If the least bound state is close to the dissociation
 threshold or moves even into the dissociative continuum, the scattering
 length and thus the interaction between the ultracold atoms are strongly
 influenced. Therefore, a small variation of the inner wall of the potential 
 can modify the interatomic interaction potential from strongly repulsive 
 to strongly attractive. This procedure is adopted in the present work 
 in order to investigate the influence of the interatomic interaction 
 potential. The scattering length is only well-defined for zero-energy 
 scattering and thus the underlying concept is in principle not applicable 
 to trapped particles with a non-vanishing zero-point energy. 
 Therefore, the scattering-length values 
 (for a given inner-wall shift) are determined for the trap-free 
 situation. In this case $a_{\rm sc}$ can be uniquely determined from the 
 analysis of the shape of the zero-energy scattering wave 
 function~\cite{cold:koeh06}.    

 Ignoring the formal problems of defining a scattering length within a 
 trap (that will be discussed in some more detail below), it is often 
 considered useful to introduce a dimension-free interaction parameter 
 $\xi$ that reflects the relative magnitude of the interaction strength 
 with respect to the confinement by the trap. If this confinement is
 approximated within the harmonic approximation, the interaction 
 parameter is naturally defined as $\xi =
 a_{\rm sc}/a_{\rm ho}$ where $a_{\rm ho}$ is the characteristic length 
 of a harmonic potential given by 
 $a_{\rm ho}=1/\sqrt{\mu\omega_{\rm{ho}}}$. For a heteronuclear 
 atom pair $\omega_{\rm{ho}}$ is again the mean harmonic frequency 
 defined in Eq.\,(\ref{eq:omegaheteronuc}). 

 Experimentally, a strong variation of the interaction strength can be
 realized with the help of magnetic Feshbach
 resonances~(MFR)~\cite{cold:loft02,cold:rega03b}. The MFR technique was also
 used to tune the interatomic interaction from strongly repulsive to strongly
 attractive in the already mentioned experiments with atoms in the optical
 lattices~\cite{cold:koeh05,cold:stoe06,cold:ospe06a}. In general, the correct
 theoretical description of a MFR requires a multi-channel scattering
 treatment which in the present case would have to incorporate also the
 optical lattice. In the analysis of the experiments described
 in~\cite{cold:koeh05,cold:stoe06,cold:ospe06a} it is, however, assumed that
 it is possible to model the MFR in an effective two-channel
 picture~\cite{cold:koeh06}. Within this model it is straightforward to
 relate the applied magnetic field to a scattering-length value (see
 Eq.~(\ref{eq:twochan}) below).

\subsection{Computational details}
\label{subsec:compdet}
 The eigenfunctions of the Hamiltonians $\oper{\rm h}_{\rm COM}$ and
 $\oper{\rm h}_{\rm REL}$ are obtained by expressing both $\psi(\vec{R}\,)$
 and $\phi(\vec{r}\,)$ as a linear combination of products of radial
 $B$-spline functions times spherical harmonics. The corresponding
 Schr\"odinger equations are solved numerically using the
 Rayleigh-Ritz-Galerkin approach~\cite{bsp:fisc90} which leads to an 
 algebraic eigenproblem.

 In general, the lattice leads to a coupling of the angular momenta. 
 Therefore, the spherical harmonics are no eigensolutions of the angular
 part. Due to the cubic trap geometry used in the experiment
 in~\cite{cold:ospe06a} and also for the present calculations, the coupling of
 different spherical harmonics is weak. In fact, the orbitals
 $\psi(\vec{R}\,)$ and $\phi(\vec{r}\,)$ describing the states relevant to 
 this work are almost converged, even if only $l=0$ is considered. 
 However, the coupling
 term $\oper{\rm W}$ in the Hamiltonian~(\ref{eq:fullham}) leads to a stronger
 angular momentum coupling. Good convergence was found in the CI
 calculation, if all spherical harmonics up to $l=3$ (and thus also $-3\leq m
 \leq +3$) were included in the calculation of the orbitals $\psi(\vec{R}\,)$
 and $\phi(\vec{r}\,)$. 

 The required number of $B$ splines and their knot
 sequence depend strongly on the behavior of the wave function
 ($\psi(\vec{R}\,)$ or $\phi(\vec{r}\,)$) that should be described. In the
 context of ultracold collisions the main interest is put on the energetically
 low-lying COM orbitals $\psi(\vec{R}\,)$ that possess a small number of 
 nodes. For the results discussed in this work, about 70 $B$ splines were 
 found to be sufficient to obtain convergence. Evidently,
 more complicated or highly anisotropic trap geometries (like double or 
 triple wells \cite{cold:schn09}) require larger expansions.

 The numerical description of the REL orbitals $\phi(\vec{r}\,)$ is more
 demanding, if one is interested in the most weakly bound states or the
 low-lying dissociative states. The BO curves of alkali-metal atom dimers support
 often a large number of bound states (for example, the \RbK system possesses 
 in the $a^3\Sigma^+$ state 
 already 32 bound states for $l=0$). The very long-ranged, 
 weakly bound states consist therefore of a highly oscillatory inner part
 (covering the so-called molecular regime and providing the orthogonality to
 all lower lying bound states) and a rather smooth long-range
 part. Correspondingly, it is practical to use two different knot sequences for
 the $B$ splines. In the present case convergence was found if 200 $B$
 splines expanded on a linear knot sequence covering the interval 
 $0\leq r\leq 20\,a_0$ are used together with about 70 $B$ splines 
 for the remaining $r$ range. The latter 70 $B$ splines are expanded on a 
 knot sequence in which the separation between the knot points increases 
 in a geometric fashion.

 Converged CI calculations were found, if they comprised configurations 
 built from about 120 REL and 60 COM orbitals. After taking symmetry into 
 account this amounts to about 1060 configurations forming the CI expansion 
 for the states of interest in this work.

\section{Results}
\label{sec:results}
 \subsection{Energy spectrum of the \RbK system}
 \label{subsec:spectrum}
 The description of an optical lattice beyond the harmonic approximation 
 is in the present work achieved by extending the Taylor expansion 
 of the $\sin^2$ potential beyond the harmonic (1st order and thus  
 quadratic) term. In principle, one should seek for convergence with 
 respect to the expansion length, but there are some practical reasons 
 why a simple convergence study as a function of the expansion length 
 causes problems. First of all, even-order expansions like the 2nd 
 order one which leads to polynomials with a degree of up to 4 (quartic 
 potential) support an infinite number of bound states with negative 
 energy, since they tend to $-\infty$ for $x$ approaching either $+\infty$ 
 or $-\infty$. However, these bound states with negative energies are 
 unphysical, since they do not exist in the case of the (original) 
 positive definite $\sin^2$ potential. The 3rd order expansion that 
 leads to polynomials up to a degree of 6 (sextic potential) supports 
 on the other hand (like all odd-order expansions) only bound states 
 with positive energy values. 

 A comparison of this sextic potential with the $\sin^2$ 
 potentials shows that the sextic potential reproduces extremely well 
 a single site of the $\sin^2$ potential and thus of the optical 
 lattice (see Fig.~\ref{fig:sketch}). Therefore, the sextic potential 
 is a good choice for the investigation of the effects of anharmonicity 
 on the bound states in a single site of an optical lattice. Evidently, the
 sextic potential cannot reproduce effects that are due to tunneling between 
 neighbor potential wells. Therefore, extended (energetically higher 
 lying) bound states in the optical lattice that are markedly affected 
 by tunneling are not well reproduced by a sextic potential. Noteworthy, 
 even in this case a simple convergence study will, however, not work. 
 For example, the 5th order and thus next odd-order expansion shows 
 a triple-well structure, but the two outer wells have a depth and 
 width that differs pronouncedly from the correct shape 
 (and the central well). This leads to completely wrongly described 
 states in these outer wells and may thus show wrong tunneling behavior 
 for the states in the middle well, especially in the case of resonant 
 tunneling. Since the present study concentrates on the anharmonicity 
 effects within a single site of an optical lattice, only the sextic 
 potential and, for comparison, the harmonic one are considered. 
 Effects that are due to tunneling between neighbor wells and thus 
 include more than single-well potentials are investigated in a separate 
 work~\cite{cold:schn09}.    

 The potential seen by the two atoms in an optical lattice contains,  
 of course, in addition to the trap potential also the interatomic 
 interaction potential that in the present case is described by 
 a Born-Oppenheimer potential curve (and appears only in the REL  
 coordinates). As is sketched in Fig.~\ref{fig:sketch}, 
 the interatomic interaction dominates the short-range part of the 
 potential and leads in the case of alkali-metal atoms to a large number of 
 bound molecular states. Since the trap potential is compared to the variation
 of the BO curve almost constant in the range of the molecular bound states, 
 especially the lower lying of these states will in practice not be influenced 
 by the optical lattice. The largest possible effect of the optical lattice 
 on the molecular bound states is expected to occur for the 
 energetically highest lying one, the least bound (lb) state. 

 Due to the large spatial extension of the trap states of typical 
 experimentally realized optical lattices these trap-induced states 
 are expected to be only weakly influenced by the molecular 
 potential. However, an immediate consequence of the existence of 
 the molecular bound states below the trap-induced ones is the nodal 
 structure at short distances that is imprinted on the wavefunctions 
 and leads to the required orthogonality of the eigenstates. The 
 energetically lowest lying and thus first trap-induced (1ti) state 
 possesses thus exactly one more node than the lb state. In the 
 experiments most closely related to the present work 
 \cite{cold:stoe06,cold:ospe06a} the transition energy between the  
 lb and the 1ti state has been measured by either rf dissociation 
 or association, respectively. This transition energy was called 
 binding energy, but it should be kept in mind that its definition 
 does not coincide with the standard definition of a 
 molecular binding energy which is given by the energy difference 
 between a molecular bound state and the (lowest) dissociation limit. 
 In the present case the existence of the optical lattice leads to 
 a discretization of the dissociation continuum and thus to an 
 additional energy shift due to the zero-point energy of the trap. 

 If the coupling $\oper{W}$ of REL and COM coordinates is ignored, 
 the energies $E_{\rm lb}$ and $E_{\rm 1ti}$ of the least bound and 
 the 1st trap-induced state, respectively, are obtained from the 
 eigenvalues of Eq.\,(\ref{eq:COMRELSchroed}) as%
  \begin{eqnarray}
    \ds
       E_{\rm lb}^{(n)}  &\equiv&  E_{\rm (1,lb)}^{(n)} 
              \:=\: \varepsilon_1^{(n)} + \epsilon_{\rm lb}^{(n)}\, 
       \label{eq:energylb}
       \\
       E_{\rm 1ti}^{(n)} &\equiv& E_{\rm (1,1ti)}^{(n)} 
              \:=\:\varepsilon_1^{(n)} + \epsilon_{\rm 1ti}^{(n)}\, 
       \label{eq:energy1ti}
  \end{eqnarray}
 where $n$ specifies the expansion length describing the optical lattice: 
 $n=2$ for a harmonic and $n=6$ for a sextic trap. In accordance with the 
 underlying assumption of an ultracold gas, the system is assumed to 
 be in its lowest state with respect to translational motion, 
 i.\,e.\ in the COM ground state with energy $\varepsilon_1$. The 
 corresponding wavefunctions are given by 
 $\Phi_{\rm lb}(\vec{R},\vec{r}\,) = \psi_1(\vec{R})\phi_{\rm lb}(\vec{r})$
 and 
 $\Phi_{\rm 1ti}(\vec{R},\vec{r}\,) = \psi_1(\vec{R})\phi_{\rm 1ti}(\vec{r})$.

 After the inclusion of the coupling of REL and COM motion the wavefunctions 
 $\Phi$ are no longer eigenstates of the Hamiltonian, but are used as 
 a basis for expanding the full wavefunctions $\Psi$ that are obtained 
 together with their energies $\eci$ by solving the Schr\"odinger 
 Eq.\,(\ref{eq:fullSchroed}). The state $\Psi$ with a dominant contribution 
 from $\Phi_{\rm lb}$ ($\Phi_{\rm 1ti}$) is then identified as least 
 bound (1st-trap-induced) state with energy $\eci_{\rm lb}^{(n)}$ 
 ($\eci_{\rm 1ti}^{(n)}$), where $n$ stands again for the order of the Taylor
 expansion of the optical-lattice potential. 

\begin{figure}[ht]
 \centering
\includegraphics[width=7.5cm,height=12cm]{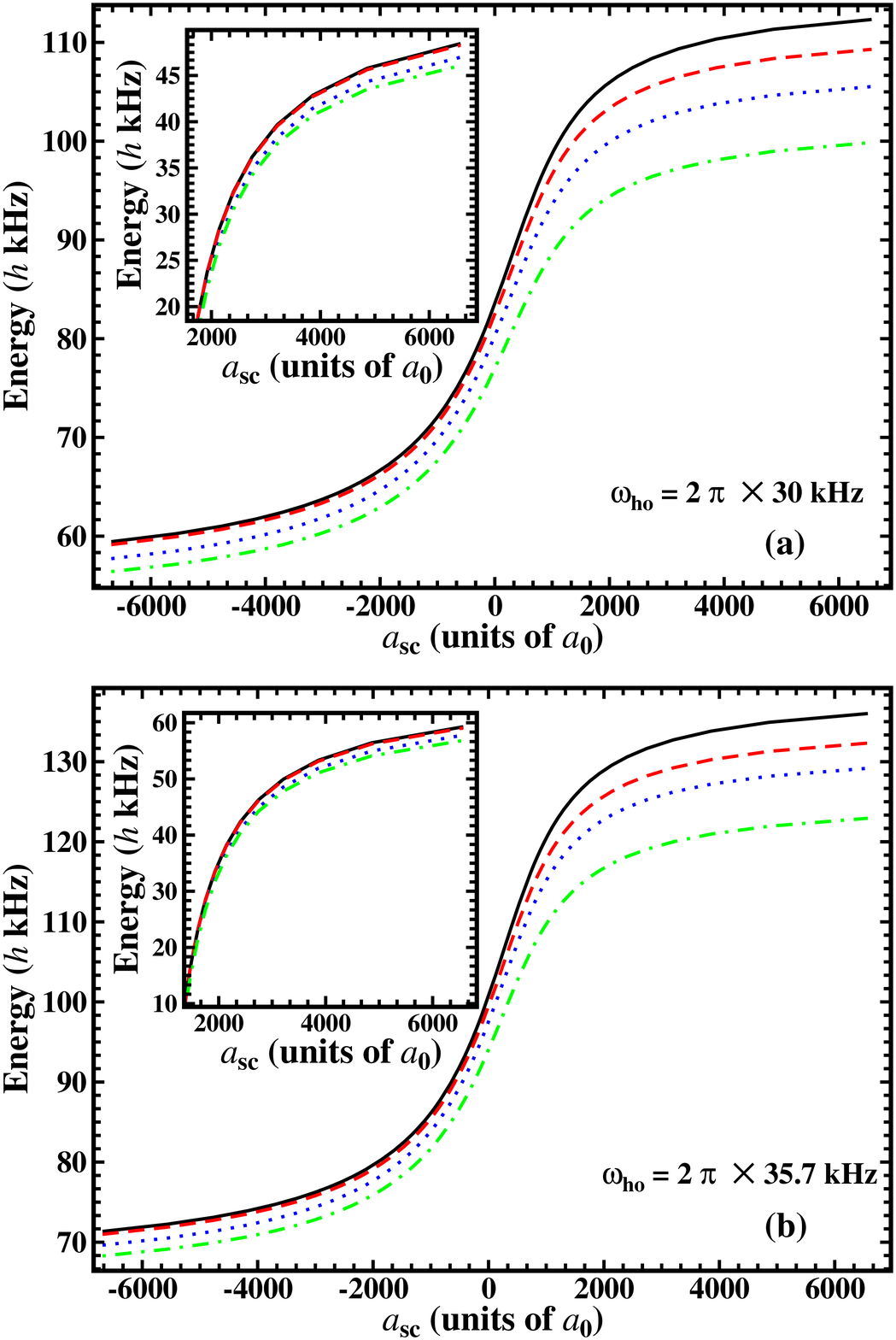}
 \caption{{\footnotesize
     (Color online)
     Energies of the 1st trap-induced (main graph) and 
     least bound (inset) state of \RbK dimers in a single site 
     of an optical lattice ($\lambda = 1030\,$nm) for the 
     potential depths 
     (a) $V_{\rm Rb}=27.5\,E_r^{\rm{Rb}}$ 
     and 
     (b) $V_{\rm Rb}=40.0\,E_r^{\rm{Rb}}$ at different levels 
     of approximation as a function 
     of the trap-free scattering length 
     (see Sec.~\ref{subsec:manofinin} for details). 
     The energies obtained with a full CI calculation for a sextic potential 
     ($\eci^{(6)}$, green chain) and a harmonic one ($\eci^{(2)}$, 
     red dashes) are compared to the corresponding sextic 
     ($E^{(6)}$, blue dots) and harmonic ($E^{(2)}$, 
     black solid) energies that are obtained, if the coupling between 
     COM and REL motion is neglected. 
     (Note, the sum $\ds \frac32(\omega_{\rm ho}+\Omega_{\rm ho})$ corresponds to
     $E^{(2)}(a_{\rm sc}=0)=100.65$ kHz.)
} }
\label{fig:RbK_27_40}
\end{figure}

 In Fig.~\ref{fig:RbK_27_40} the energies of the least bound and the 
 1st trap-induced state are shown for \RbK as a function 
 of the trap-free scattering length (see Sec.~\ref{subsec:manofinin}) 
 for different levels of approximation ranging from the separable 
 harmonic one to the fully coupled sextic solution. %
 The trapping parameters were chosen in accordance with the corresponding 
 experiment~\cite{cold:ospe06a}. Clearly, the energies for the different 
 approximations differ most for large positive scattering lengths and 
 thus in the case of a strong repulsive interaction between the atoms. 

 A comparison of the results for the least bound and the first trap-induced
 states reveals that the energy of the former is almost unaffected by the
 anharmonicity of the trap and COM-REL coupling. Although this state is at
 least in the strongly repulsive part of the spectrum long ranged, it "feels"
 the anharmonic form of the trapping potential very weakly. This state
 remains thus sufficiently deeply localized in the trap potential and the
 harmonic approximation works still reasonably well; even for the rather
 strong repulsive interaction expressed by the scattering length 
 $a_{\rm sc}\approx 6500\,a_0$. The energy change of the first trap-induced
 state due to the anharmonicity and REL-COM coupling is on the other hand
 much more pronounced. This change is thus predominantly defining the
 modification of the binding energy due to the trap.

 If the results for different levels of approximation are compared with each 
 other, a clear ordering is visible. Independent of the scattering length and 
 thus the interaction strength as well as its type (repulsive or attractive) 
 the uncoupled harmonic energy $E^{(2)}$ is lowered, if the COM-REL coupling 
 is included ($\eci^{(2)}$). Note, this coupling exists even within the harmonic 
 approximation for a heteronuclear diatomic molecule like RbK, since the 
 two atoms possess different masses and polarizabilities and experience 
 therefore different trap potentials. As a consequence, COM and REL motions 
 do not separate. Only for diatomic systems made from two atoms in the same
 electronic state (or in some accidental situation) this coupling of 
 COM and REL motion vanishes within the harmonic approximation.   

 An even larger reduction of the energy is observed, if only the
 anharmonicity is considered as is reflected by $E^{(6)}$ in which the 
 coupling of COM and REL motion is ignored. For the considered system 
 the effect of anharmonicity is thus a larger correction to the separable 
 harmonic approximation than the one caused by the coupling of COM and REL 
 motion. A further energy reduction is found, if both effects are considered 
 which leads to $\eci^{(6)}$. 
 Interestingly, the energy reduction indicated by $\eci^{(6)}$ is larger 
 than the sum of the energy reductions obtained separately for $\eci^{(2)}$ 
 and $E^{(6)}$. The coupling of COM and REL motion is thus enhanced, if 
 the more realistic sextic potential is considered instead of the harmonic 
 one.   
   
 In order to quantitatively describe the different effects of the 
 trapping potential the energy differences %
  \begin{eqnarray}
    \ds
      \Delta_{\rm geom} &=& E_{i}^{(2)} - E_{i}^{(6)}
      \label{eq:geomd}
      \\
      \Delta_{\rm coup}^{(n)} &=&  E_{i}^{(n)} - \eci_{i}^{(n)}
      \label{eq:coupld}
      \\
      \Delta_{\rm tot} &=&  E_{i}^{(2)} - \eci_{i}^{(6)} 
                        = \Delta_{\rm geom} + \Delta_{\rm coup}^{(6)} 
      \label{eq:totald}
  \end{eqnarray}   
 may be introduced, where $i=\{{\rm lb,1ti}\}$. $\Delta_{\rm geom}$ characterizes the 
 effect of anharmonicity of the optical lattice based on the uncoupled solutions. 
 $\Delta_{\rm coup}^{(n)}$ is a measure of the coupling between 
 COM and REL motion within the harmonic ($n=2$) or sextic ($n=6$) 
 potential. Finally, $\Delta_{\rm tot}$ specifies the energy difference 
 between the simple harmonic approximation (in which the coupling of 
 COM and REL motion is ignored) and the full solution of two atoms 
 in a single site of an optical lattice (within the sextic approximation). 

 As is evident from Fig.~\ref{fig:RbK_27_40}, the effect is largest for the strongly
 repulsive regime. This is due to a rise of the energy level to the region of 
 higher anharmonicity of the trapping
 potential. Moreover, the state in this point is also long-ranged due to the
 strong repulsive interaction. Numerical values of the differences 
 $\Delta$~(\ref{eq:geomd}-\ref{eq:totald}) for \RbK, the
 experimental trap parameters in \cite{cold:ospe06a}, and $a_{\rm sc}=6500\,a_0$
 are given in Table~\ref{tab:deltas}.
\begin{table}
  \caption{The effect of the trapping potential on the energy 
           spectrum of the 1st trap-induced state of \RbK for 
           the trapping parameters of the experiment reported 
           in~\cite{cold:ospe06a}. The $\Delta$ values defined by
           Eqs.~(\ref{eq:geomd})-(\ref{eq:totald}) are given
           in units of $h^{-1}/\text{kHz}$ and calculated at
           $a_{\rm sc}=6500\,a_0$.} 
  \begin{ruledtabular}
     \begin{tabular}{ccccc}
        $V_{\rm Rb}(E_r^{\rm{Rb}})$ & $\Delta_{\rm geom}$ & $\Delta_{\rm coup}^{(2)}$ &
         $\Delta_{\rm coup}^{(6)}$ & $\Delta_{\rm tot}$ \\
         \vspace{-0.3cm}
         \;& \;& \;& \; & \; \\
        \hline
         \vspace{-0.3cm}
         \;& \;& \;& \;& \;  \\
         27.5  \;  & 6.797 & 3.022  & 5.665  & 12.462 \\
         40.0  \;  & 6.828 & 3.689  & 6.243  & 13.071 \\
     \end{tabular}
  \end{ruledtabular}
  \label{tab:deltas}
\end{table}

 For the considered system the value $\Delta_{\rm tot}$ and thus the total
 energy between the uncoupled and the coupled harmonic approximation amounts
 to about 13\,kHz for $a_{\rm sc}=6500a_0$. An effect of this size should be
 visible in the experiment in~\cite{cold:ospe06a} with a claimed resolution of
 1.7\,kHz but would not be resolvable with a ten times worse resolution as it
 occurs for a ten times shorter rf-pulse as was used, e.\,g.,
 in~\cite{cold:stoe06}.

 In an analysis of the influence of the interaction strength as is performed 
 in this work it is important to stay within the restrictions of a single-site 
 model. The parameter variation has to avoid situations in which tunneling 
 or even over-the-barrier transfer of atoms between different sites of a 
 physical optical lattice can occur, since this range is clearly not 
 adequately described with a harmonic or sextic potential with infinite 
 walls. For example, for a very large positive scattering 
 length the large repulsive interaction shifts the lowest-lying atom-pair 
 state (1st trap-induced state) above the barrier of a true optical lattice. 
 While this physical lattice would not support any bound states, the 
 harmonic or sextic potentials would still possess an infinite number 
 of them. As is discussed in Sec.\,\ref{sec:wf}, it was always checked 
 that the wave functions remain well localized within the boundaries 
 of a single site of the optical lattice for the parameters used in 
 this study.       

 \subsection{Other systems}
 \label{subsec:othsys}
\begin{figure}[ht]
 \centering
\includegraphics[width=8cm,height=16cm]{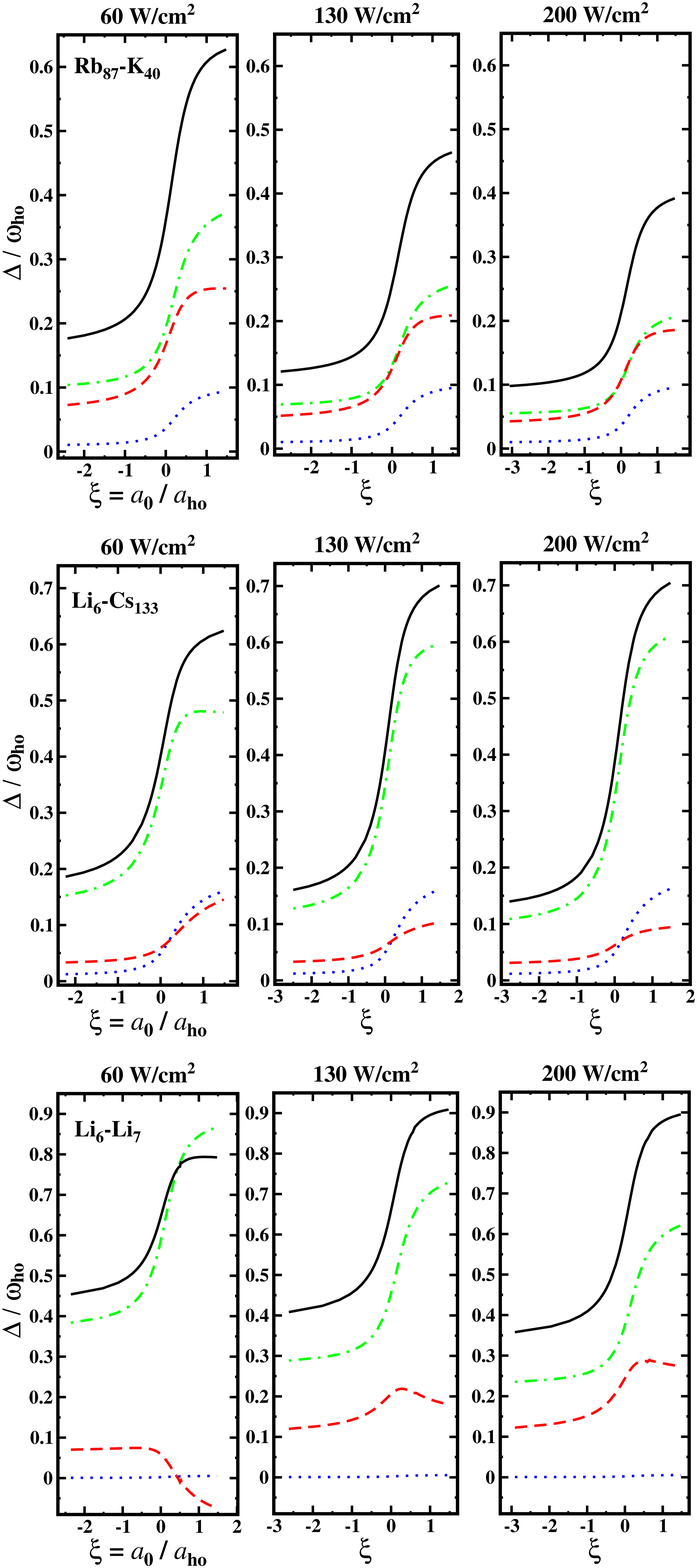}
 \caption{{\footnotesize
     (Color online)
     The energy differences  
     [see Eqs.~(\ref{eq:geomd}-\ref{eq:totald})]
     $\Delta_{\rm coup}^{(2)}$ (blue dots),
     $\Delta_{\rm coup}^{(6)}$ (red dashes),
     $\Delta_{\rm geom}$ (green chain),
     $\Delta_{\rm tot}$ (black solid) 
     (in multiples of $\omega_{\rm ho}$, both in 
      atomic units) 
     for different alkali metal dimers and intensities
     of the lattice laser (as specified in the 
     graphs). The wavelength of the trap laser  
     is 1030\,nm. (The laser intensity 200 W/cm$^2$ 
     corresponds to a $30\,E_r^{\rm{Rb}}$ lattice 
     depth for the \RbK dimer.) 
} }
\label{fig:deltas}
\end{figure}
 In order to obtain a more complete picture of the anharmonicity and the coupling
 effects other systems may be analyzed. Besides the already considered \RbK pair
 (example of large masses and polarizabilities) other experimentally relevant
 alkali metal dimers like \LiCs (small mass and polarizability of $^6$Li and for $^{133}$Cs
 both characteristics are large) and \LiLi (small masses and polarizabilities
 of both elements) are investigated.

 Figure~\ref{fig:deltas} shows the differences
 $\Delta$~(\ref{eq:geomd}-\ref{eq:totald}) as a function of the scaled interaction 
 parameter $\xi$ (see Sec.~\ref{subsec:manofinin}) for different lattice depths 
 obtained by the laser intensity variation for the three mentioned systems. As
 is evident from Fig.~\ref{fig:deltas}, the harmonic coupling difference
 $\Delta_{\rm coup}^{(2)}$ is not
 influenced by the lattice depth, because the coupling depends only on
 the polarizabilities and the masses and is of the form
 $(\mu_2\alpha_1-\mu_1\alpha_2)$. Therefore, $\Delta_{\rm coup}^{(2)}$ is largest 
 for \LiCs and smallest for \LiLi as is clear from  Fig.~\ref{fig:deltas}. 
 Beyond the harmonic approximation the mass, polarizability, laser
 intensity, and $k$-vector dependence are mathematically non-trivial in the 
 framework of the present approach. As a result, the different $\Delta$ values
 have a behavior which is difficult to predict {\it a priori}. For example, while the total
 difference $\Delta_{\rm tot}$ decreases with the laser intensity for \RbK and
 increases for the other two systems, the values $\Delta_6$ and 
 $\Delta_{\rm geom}$ change their behavior not only with the laser intensity
 but also when going from one dimer to the other. Most noteworthy, the $\Delta$ values
 for \LiLi are not smaller than for the other pairs although this system is
 almost homonuclear. 

 Another peculiar feature of the \LiLi dimer compared to the other considered
 ones is the occurrence of negative values for $\Delta_{\rm coup}^{(6)}$ in
 the case of large positive values of $\xi$ and the laser intensity of 
 60\,W\,cm$^{-2}$. This leads to a smaller value of $\Delta_{\rm tot}$ 
 compared to $\Delta_{\rm geom}$ for these parameters. Clearly, the
 conclusions obtained for the generic \RbK system are not always 
 transferable to other alkali metal dimers.  

 \subsection{Wave-function analysis}
 \label{sec:wf}
 \subsubsection{Radial pair densities}
 \label{subsec:raddens}
 \begin{figure}[ht]
  \centering
 \includegraphics[width=7.7cm,height=16.8cm]{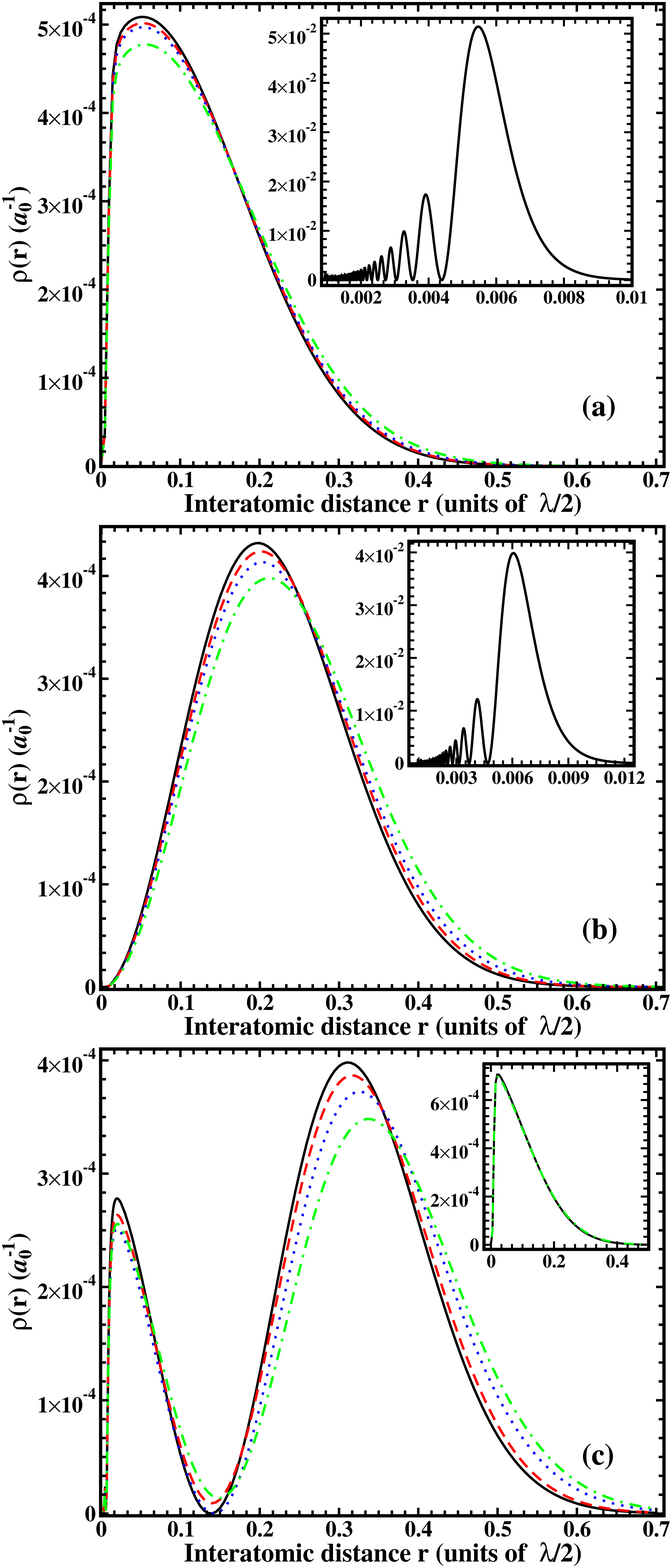}
  \caption{{\footnotesize
      (Color online)
      Radial pair densities of the 1st trap-induced state of the \RbK system 
      in a 3D cubic lattice of $40\,E_r^{\rm{Rb}}$ depth and the laser 
      wavelength $\lambda$ of 1030\,nm in the uncoupled harmonic (black
      solid), the harmonic with coupling (red dashes), the uncoupled sextic  
      (blue dots), and the sextic with coupling (green chain) approximations 
      for the three interaction regimes: a) strongly attractive, 
      b) almost zero interaction, c) strongly repulsive.
      (The insets show the densities for the least bound-state.)
 } }
 \label{fig:1D_WFD_RbK40_ascMZP}
 \end{figure}
 An alternative analysis of the anharmonicity and COM-REL coupling effects is 
 possible from the wave functions of the first trap induced and the least
 bound state. Since the probability density for finding a two-particle
 separation to lie in between $r$ and $r+dr$ is determined by the radial pair density  
 \begin{equation}
   \rho(r) = \int\!\!\int\!\!\int\!\!\int\!\!\int \:  
           |\chi_i(\vec{R},\vec{r})|^2 \: 
           {\rm d}V_{R} \:
           r^2 \: {\rm d}\Omega_r
           \quad ,
  \label{eq:wfdens}
 \end{equation}
 it is convenient to discuss radial pair densities instead of the wave functions. In
 Eq.~(\ref{eq:wfdens}) the function $\ket{\chi_i}$ stands for 
 $\ket{\Psi_i}$ or $\ket{\Phi_i}$ depending on the considered approximation,
 ${\rm d}V_{R}$ is the COM volume element, and $\Omega_r$ is the angular part
 of the REL motion coordinates.

 The energy spectrum of the first trap induced and the least bound states for
 the wide range of the interaction regimes was presented in
 Fig.~\ref{fig:RbK_27_40}. However, the three asymptotic interaction
 situations, namely, strong attraction ($a_{\rm sc}\rightarrow -\infty$), the
 almost zero interaction ($a_{\rm sc}\rightarrow 0$) and the strong repulsion
 ($a_{\rm sc}\rightarrow +\infty$) are found sufficient for the wave function
 analysis. Figure~\ref{fig:1D_WFD_RbK40_ascMZP} shows the radial pair
 densities at
 the different levels of approximation for the three interaction regimes.
 As is evident from Fig.~\ref{fig:1D_WFD_RbK40_ascMZP}, a large
 attractive interaction leads to a very confined function for the first
 trap-induced bound state while a large repulsive interaction does not only
 result in a node but also in a shift of the outermost lobe to larger
 interatomic distances. This shift is counteracted by the confinement of the
 trap. Remind, for the trap-free situation in the strongly repulsive regime the
 wave function crosses the internuclear axis exactly at the value of the scattering
 length. Evidently, the behavior of the density for almost zero interaction is
 only determined by the trap.
 
 As is apparent from Fig.~\ref{fig:1D_WFD_RbK40_ascMZP} the inclusion of
 the anharmonicity and the coupling leads to more extended pair densities. For all
 interaction regimes the following behavior is found. The effect is smallest 
 for the harmonic coupling correction. A larger effect is found for the sextic 
 non-coupled case which is strengthened by the sextic coupling. 
 Effectively, the particles experience a
 more extended trap, if a more complete description of the problem is
 achieved. Such an effect is expected for the harmonic-to-sextic uncoupled
 description, since the harmonic trap is tighter than the sextic one as is
 evident from the sketch in Fig.~\ref{fig:sketch}. While this is expected for
 the inclusion of the anharmonicity, this is not immediately clear for the 
 coupling.

 The least bound state in the strongly repulsive regime is very long-ranged
 as the inset of Fig.~\ref{fig:1D_WFD_RbK40_ascMZP}(c) shows. This distance is
 almost comparable with the extension of the first trap-induced state for the
 strongly attractive regime. Nevertheless, the influence of the anharmonicity
 and the coupling on the least bound state is almost absent, 
 because the state is energetically deeply bound and therefore does almost 
 not probe the anharmonicity of the lattice. 

 In the considered parameter ranges the radial pair densities approach zero 
 clearly before the interatomic distance reaches the boundary of a single 
 lattice site, i.\,e.\ for $r<\lambda/2$, as can be seen from 
 Fig.~\ref{fig:1D_WFD_RbK40_ascMZP}. Therefore, tunneling or a
 distribution of the dimer over more than a single lattice site does not 
 occur and the present single-site model is applicable. Furthermore, 
 effects of the artificial infinite walls of the harmonic or sextic potentials 
 should not be a problem. However, the radial pair densities provide only 
 an indication for the applicability of the single-site approximation, since 
 it is still possible that the dimer as a whole may be distributed over 
 more than one site. This can only be excluded form an analysis of the 
 total wave function including the COM motion as is done in the next section.         
 
 For the present type of the coupled problem analysis the plots of the radial
 pair densities can be used as a good check of the validity of the pseudopotential
 approximation. Varying the value of the scattering length in the
 pseudopotential approach it is possible to match the function of the full
 solution at the long range distance. Such an approach is comparable to the
 energy-dependent concept developed for two identical particles in a harmonic
 trap~\cite{cold:blum02} and will be discussed in a separate work. Before 
 entering such an analysis, it may be remarked that the radial pair densities 
 as shown in Fig.~\ref{fig:1D_WFD_RbK40_ascMZP} are also of interest for 
 the investigation of the validity of a pseudopotential approximation for 
 the interatomic interaction. In fact, it may even be used in order to 
 attempt to obtain an improved pseudopotential describing atomic pairs 
 in an optical lattice. A corresponding study is presently underway.

 \subsubsection{Wave function in absolute coordinates}
 \label{subsec:wfabsolute}
 It is instructive to analyze the full wave function or corresponding particle
 density also in absolute coordinates
 of the laboratory space (ABS). They supply the complete information about
 the dimer and provide pictures of the COM and REL motion simultaneously. 
 This is evidently not the case for the radial pair density 
 (Fig.~\ref{fig:1D_WFD_RbK40_ascMZP}) that is averaged over the COM motion and
 thus does not reveal whether the pair as a whole moves through the lattice. 
 Note, while the radial pair density provides nevertheless a rather easy to 
 interpret picture of the underlying physics, this is far less the case for 
 its angular part. The reason is that the anisotropic (egg-box like) shape 
 of the (cubic) optical lattice does not trivially show up in the REL
 coordinate system. An analysis in the lab frame is, however, also
 non-trivial, since 
 the functions depend on six spatial coordinates. Use of the cubic symmetry 
 reduces the size of the symmetry non-equivalent space, but it is still 
 impractical to consider the complete multidimensional function. Instead, 
 some insight may be gained from selected cuts. Although a number of cuts was 
 analyzed in this work, only the results for cuts along the $x$ coordinate 
 of both atoms and thus for $y_i=z_i=0$ (for both particles $i$) are 
 shown and discussed.  

 \begin{figure}[ht]
   \centering
   \includegraphics[width=8.5cm,height=12cm]{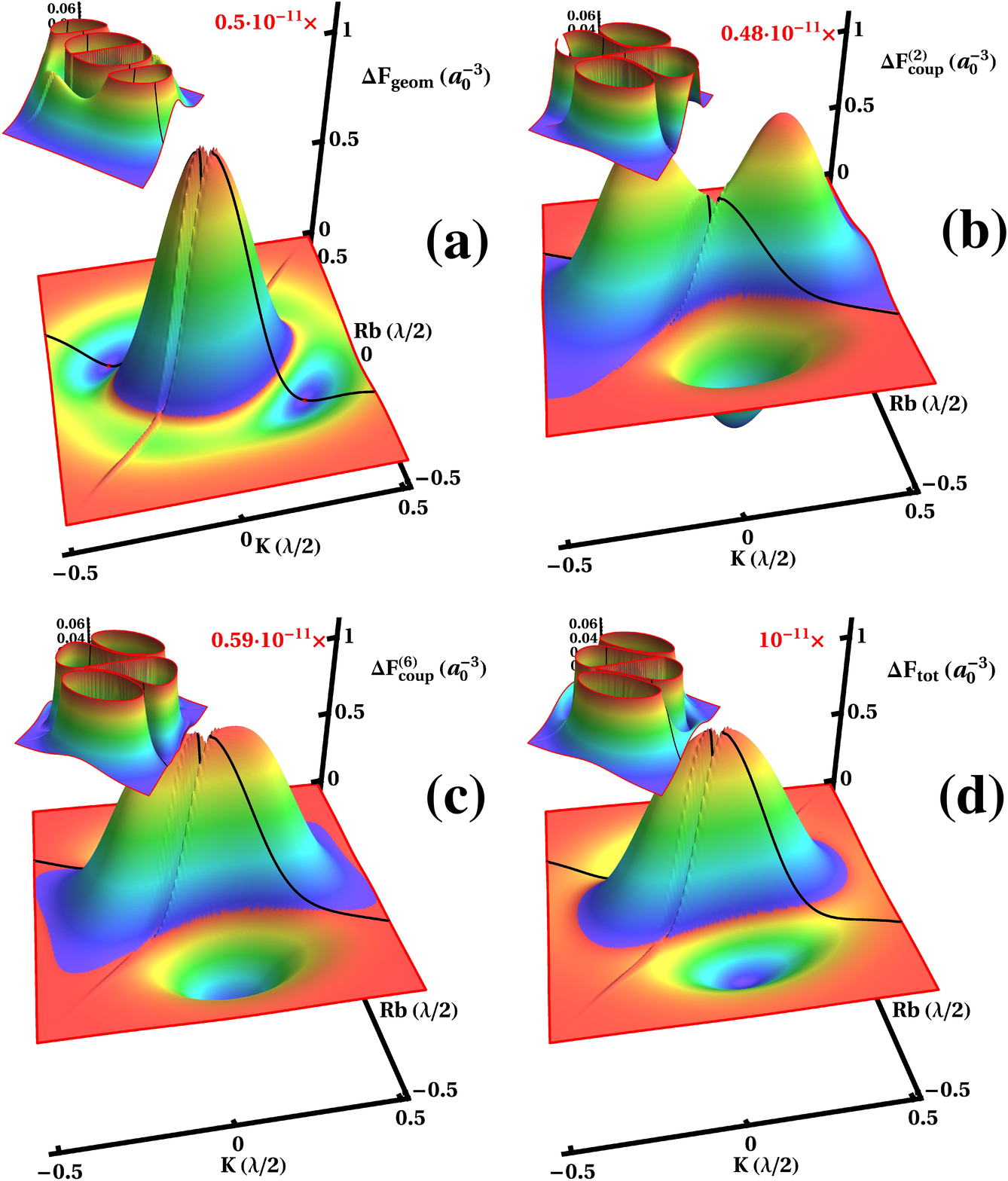}
   \caption{{\footnotesize
           (Color online)
           Cuts of $\Delta F$ (defined in 
           Eqs.~(\ref{eq:geomdwf}-\ref{eq:totaldwf}))
           along the $x$ direction 
           ($y_{\rm Rb}=y_{\rm K}=z_{\rm Rb}=z_{\rm K}=0$)  
           for the 1st trap-induced state and almost   
           non-interacting \RbK atoms in a 3D cubic lattice 
           ($40\,E_r^{\rm{Rb}}$, $\lambda$ = 1030\,nm):  
           a) $\Delta F_{\rm geom}(x_{\rm Rb},x_{\rm K})$
           b) $\Delta F_{\rm coup}^{(2)}(x_{\rm Rb},x_{\rm K})$
           c) $\Delta F_{\rm coup}^{(6)}(x_{\rm Rb},x_{\rm K})$
           d) $\Delta F_{\rm tot}(x_{\rm Rb},x_{\rm K})$. 
           The differences $\Delta F$ are given in atomic 
           units, downscaled by the corresponding factors 
           given in red. 
           The black lines indicate the COM axis 
           ($x_{\rm Rb}=-\mu_{\rm K}/\mu_{\rm Rb}\, x_{\rm K}$).
           The insets show $|\Delta F|$ (on an enlarged scale). 
        }}
     \label{fig:cutZ}
 \end{figure}
 In order to quantitatively describe the different effects of the trapping
 potential it is again useful to consider not the wave functions at different 
 level of approximation themself, but their respective differences. 
 Similarly to the energy differences defined in 
 Eqs.~(\ref{eq:geomd}-\ref{eq:totald}), the wave-function differences 
  \begin{eqnarray}
    \ds
      \Delta F_{\rm geom}(\vec{r}_1,\vec{r}_2) &=&
      \Phi_{i}^{(2)}(\vec{r}_1,\vec{r}_2) - 
      \Phi_{i}^{(6)}(\vec{r}_1,\vec{r}_2)
      \label{eq:geomdwf}
      \\
      \Delta F_{\rm coup}^{(n)}(\vec{r}_1,\vec{r}_2) &=&
      \Phi_{i}^{(n)}(\vec{r}_1,\vec{r}_2) -
      \Psi_{i}^{(n)}(\vec{r}_1,\vec{r}_2)
      \label{eq:coupldwf}
      \\
      \Delta F_{\rm tot}(\vec{r}_1,\vec{r}_2) &=&
      \Phi_{i}^{(2)}(\vec{r}_1,\vec{r}_2) -
      \Psi_{i}^{(6)}(\vec{r}_1,\vec{r}_2) 
      \label{eq:totaldwf}
  \end{eqnarray}
 may be introduced. Cuts through these difference functions $\Delta F$ are 
 shown in Fig.~\ref{fig:cutZ} for the first trap-induced state and the almost 
 non-interacting case ($a_{\rm sc}\approx 0$). The sign convention used 
 in Eqs.~(\ref{eq:geomdwf}-\ref{eq:totaldwf}) means that positive maxima in 
 Fig.~\ref{fig:cutZ} correspond to the case that the wave functions in lower 
 order of approximation have a larger amplitude than those in the higher one. 
 (This choice is, of course, arbitrary and basically motivated by the fact 
 that it leads to positive maxima in the center of the plots which is more 
 suitable for optical reasons.) 

 The diagonal 
 $x_{\rm Rb}=x_{\rm K}$ defines the REL coordinate axis. The wavefunctions 
 and therefore also their differences $\Delta F$ are strictly zero along 
 the REL axis, since the molecular interaction potential rises exponentially 
 to infinity for $r \rightarrow 0$. Note, even for $a_{\rm sc}=0$ the 
 atoms interact in the present approach, since the scattering length 
 characterizes only the effective long-range interaction. In the case 
 of the often adopted $\delta$-type pseudopotential description the 
 interaction vanishes completely for $a_{\rm sc}=0$ and the wavefunction 
 does not vanish around $r=0$. Slightly away from the REL axis the 
 wavefunction shows rapid oscillations due to the nodal structure that 
 is again a consequence of the realistic interatomic interaction 
 potential used in the present work. They are, however, not resolved 
 in Fig~\ref{fig:cutZ}, as these oscillations occur in a very small 
 ($\ds \sim 10^{-3}\,\lambda/2$) $r$ range compared to the one displayed. 

 The COM axis is defined by 
 $x_{\rm Rb}=-\mu_{\rm K}/\mu_{\rm Rb}\, x_{\rm K}$. Since \RbK is 
 heteronuclear, the COM axis is rotated from the $x_{\rm Rb}=-x_{\rm K}$ diagonal and is 
 thus for better readability explicitly indicated in the graphs. 
 Another consequence of the heteronuclear character is the elliptical 
 shape of the figures contours that would be circular in the case of 
 a homonuclear system.

 Figure \ref{fig:cutZ}(a) characterizes the geometrical effect of the 
 anharmonicity of an optical lattice. Effectively, the sextic trap is more
 extended than the harmonic one. This leads to the decrease of the density at
 the center of the potential and an increased probability at the potential
 edges. Therefore, the probability to find Rb and K atoms at a larger distance
 from the center of the optical lattice is higher for an anharmonic trap 
 compared with a harmonic one. Note, this probability redistribution is not
 homogeneous. For example, for $\ds x_{\rm K}\approx 0.3\,\lambda/2$
 and $\ds x_{\rm Rb}\approx -0.1\,\lambda/2$ a pronounced minimum of the
 function $\Delta F_{\rm geom}$ exists, as is evident
 from Fig.~\ref{fig:cutZ}(a). This behavior in the ABS space is a direct consequence of the different
 COM and REL motion trapping depths. The COM of the system is more
 confined. Hence, the density shift in the direction of the COM axis is larger
 than for the REL one, as is better seen in the inset of
 Fig.~\ref{fig:cutZ}(a). In the inset one notices also that there exist 
 small minima at the places where both atoms are close together 
 ($r\approx 0$), but away from the center of the lattice site. In general,  
 $\Delta F_{\rm geom}$ is rather symmetric with respect to the COM and REL 
 axes.

 Figure \ref{fig:cutZ}(b) shows the effect of the coupling of COM and 
 REL motion within the harmonic approximation. 
 For its understanding it is important to keep in mind that the definition 
 of coupling between the different degrees of freedom depends on the adopted 
 coordinate system. In the present work it is defined by the Hamiltonian 
 in Eq.~(\ref{eq:fullham}) and thus the coupling of COM and REL coordinates. 
 While this is a natural choice for discussions of, e.\,g., the radial 
 pair density, its meaning is less transparent for a discussion of 
 wave functions in ABS coordinates of the two atoms. An 
 evident example is the case of two truly non-interacting atoms in a 
 harmonic trap. Even for a heteronuclear atom pair the problem separates 
 in ABS coordinates, as was already mentioned in Sec.~\ref{subsec:hamilt}. 
 However, treating this system in COM and REL coordinates the coupling 
 term $\hat{W}$ in Eq.~(\ref{eq:fullham}) and thus also the difference 
 $\Delta F_{\rm coup}^{(2)}$ is non-zero, but the latter reflects the 
 non-separability due to the adopted coordinate system. 

 A comparison of $\Delta F_{\rm coup}^{(2)}$ shown in Fig.~\ref{fig:cutZ}(b) 
 with the one obtained from the analytically known harmonic solutions in 
 either ABS or REL and COM coordinates for truly non-interacting particles 
 (see, e.\,g., \cite{cold:dawy67}) confirms that the structures in 
 Fig.~\ref{fig:cutZ}(b) for the \RbK dimer with 
 the long-range interaction being tuned to be almost vanishing are 
 similar. In contrast to the case of the geometry effect visible in  
 Fig.~\ref{fig:cutZ}(a) the maxima and minima in Fig.~\ref{fig:cutZ}(b) 
 have a more similar magnitude (the maxima being about 16\,\% larger 
 in absolute value than the minima) and, clearly, they also originate 
 from the different effects discussed above. Furthermore, they are located away 
 from the center of the optical-lattice site. In fact, they are 
 found, if one of the two atoms is located closely to the center and the 
 other one is separated by about the most likely separation (about 
 $0.2\,\lambda/2$, see Fig.~\ref{fig:1D_WFD_RbK40_ascMZP} (b)). The 
 maxima (minima) are connected with the lighter K (heavier Rb) atom 
 being close to the center. The anti-clockwise rotation of the maxima 
 and minima around the origin is due to the heteronuclear 
 character and reflects the coupling term in ABS 
 coordinates, namely, 
 $\Delta F_{\rm coup}^{(2)}\sim e^{-\gamma x_{\rm K} x_{\rm Rb}}$ 
 (where $\gamma$ is some constant). Also the different widths of 
 the maxima and minima is a consequence of the heteronuclear character 
 of \RbK. While the off-centered minima 
 of $\Delta F_{\rm geom}$ are centered on the COM axis, the COM 
 axis appear to separate the minima and maxima of 
 $\Delta F_{\rm coup}^{(2)}$, although it does not define a strict 
 nodal plane.     

 The sextic coupling effect presented in Fig.~\ref{fig:cutZ}(c) is 
 similar to the harmonic one in Fig.~\ref{fig:cutZ}(b). However, the 
 two maxima are now almost connected (if there were not the strict 
 node on the REL axis) and form more a kind of plateau. The 
 minima are less pronounced and as a consequence, the absolute 
 values of the maxima are now about $40\%$ larger than the ones 
 of the minima.  

 Figure~\ref{fig:cutZ}(d) presents the complete effect of anharmonicity and
 coupling of the optical lattice. Compared to the previously discussed 
 $\Delta F$ functions the shown $\Delta F_{\rm tot}$ is in shape most 
 similar to $\Delta F_{\rm coup}^{(6)}$ in Fig.~\ref{fig:cutZ}(c). However, 
 the two maxima at the corners of the plateau appear now to be merged with the 
 central peak due to the new scale (and are basically only separated by the 
 node along the REL axis). 
 As a consequence, the density of the exact sextic solution is reduced at 
 the center of the lattice compared with the uncoupled harmonic 
 approximation. In fact, as is evident from  the equality 
 $\Delta F_{\rm tot}= \Delta F_{\rm geom}+\Delta F_{\rm coup}^{(6)}$ 
 (compare Eq.~(\ref{eq:totald})), the merging of the two maxima is 
 simply a consequence of the superposition of the structures of 
 $\Delta F_{\rm coup}^{(6)}$ and $\Delta F_{\rm geom}$. Note the 
 correspondingly almost by a factor 2 larger amplitude of 
 $\Delta F_{\rm tot}$ compared to the other wavefunction 
 differences. Since the minima of $\Delta F_{\rm geom}$ and 
 $\Delta F_{\rm coup}^{(6)}$ appear at rather different places, 
 their relative importance diminishes in comparison to the 
 maxima. This leads to an about 66\,\% larger absolute value 
 of the maxima compared to the minima in the case of 
 $\Delta F_{\rm tot}$. However, the additivity leads to an effective 
 broadening of the minima of $\Delta F_{\rm tot}$ in direction of 
 the COM axis compared to the minima found for 
 $\Delta F_{\rm coup}^{(6)}$.

 To conclude the almost non-interacting case, the optical lattice is in 
 the coupled sextic description effectively more extended than in the 
 uncoupled harmonic one. 
 As a consequence of this anharmonicity the wave-function amplitude at the center 
 of the lattice site decreases and is redistributed to the edges of the potential. 
 As a consequence of the coupling, the decrease of the wave-function 
 amplitude stretches further out along a diagonal in between the 
 COM and REL axes close to the axis defined by the Rb atom being  
 located at the center of the lattice site ($x_{\rm Rb}=0$). On the 
 other hand, the coupling leads also to minima (increased amplitude) 
 along a diagonal between the COM and REL axes, but close to the 
 $x_{\rm K}=0$ axis. As a consequence of the heteronuclear character of 
 the \RbK dimer, the two diagonals are rotated with respect to the two  
 corresponding $x=0$ axes.   

   \begin{figure}[ht]
     \centering
     \includegraphics[width=8.5cm,height=12cm]{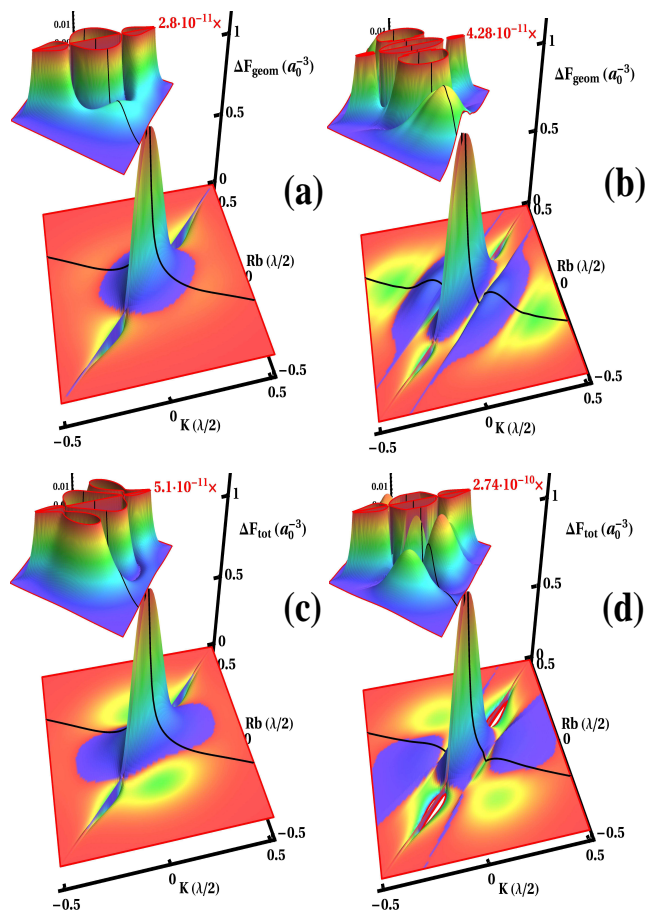}
      \caption{{\footnotesize
          (Color online)
          Cuts through the differences $\Delta F$ as in 
          Fig.~\ref{fig:cutZ}, but
          a) $\Delta F_{\rm geom}(x_{\rm Rb},x_{\rm K})$ and
          c) $\Delta F_{\rm tot}(x_{\rm Rb},x_{\rm K})$ for  
             strongly attractively interacting particles 
             ($a_{\rm sc}=-6600\;a_0$);
          b) $\Delta F_{\rm geom}(x_{\rm Rb},x_{\rm K})$ and 
          d) $\Delta F_{\rm tot}(x_{\rm Rb},x_{\rm K})$ for 
             strongly repulsively interacting particles 
             ($a_{\rm sc}=+6600\;a_0$).      
     } }
    \label{fig:cutPM}
    \end{figure}
 Figure~\ref{fig:cutPM} characterizes the anharmonicity and coupling effect in
 the strongly interacting regimes. The most evident difference between the 
 (almost) non-interacting case and the strongly interacting situations is 
 the pronounced squeezing of the central peak along the COM axis. This 
 is easily seen by comparing the geometry effect characterized by $\Delta
 F_{\rm geom}$ in Figs.~\ref{fig:cutPM}(a) and (b) with 
 Fig.~\ref{fig:cutZ}(a). Connected with this squeezing is an increase of the 
 maxima by a factor of almost 6 (strong repulsion) or more than 8 (strong 
 attraction). The additional lobe occurring for strongly repulsive interaction 
 at large distances leads to two further maxima on the COM axis (one for 
 a positive and one for a negative value of $x_{\rm K}$) for 
 $\Delta F_{\rm geom}$ (Fig.~\ref{fig:cutPM}(b)), indicating a corresponding 
 difference between the uncoupled harmonic and sextic solutions that occurs 
 also at the outer lobes. The mimima on the COM 
 axis are in this case shifted to larger distances from the REL axis. 
 This is not the case for a strong attractive interaction, but there the 
 amplitude of the minimum is even smaller than in the attractive case  
 where it is already of less relative importance compared to the central 
 maxima than in the non-interacting case. On the other hand, the minima 
 on the REL axis that had been very weak compared to the ones on the 
 COM axis for the non-interacting case are in the strongly interacting cases 
 much more pronounced, but also squeezed into a narrow regime close to 
 the REL axis.  

 As for the almost non-interacting case (Fig.~\ref{fig:cutZ}(d)), the total 
 differences $\Delta F_{\rm tot}$ for the strongly interacting cases 
 (Figs.~\ref{fig:cutPM}(c) and (d)) differ from their 
 $\Delta F_{\rm geom}$ counterparts by the occurrence of two minima along 
 diagonals between the COM and REL axes for $x_{\rm K} \approx 0$. 
 While the also coupling-induced maxima for $x_{\rm Rb} \approx 0$ lead 
 for the non-interacting case to a broad central peak, they appear in 
 the strongly interacting case as shoulders. The reason is the massive 
 squeezing of the central peak already discussed for $\Delta F_{\rm geom}$. 
 The additional maxima along the COM axis in the case of strong repulsion
 lead to a rather structured difference surface $\Delta F_{\rm tot}$ in this 
 case. Another interesting effect visible from Fig.~\ref{fig:cutPM}(d) is 
 the enormous increase of the central maximum when comparing 
 $\Delta F_{\rm tot}$ with $\Delta F_{\rm geom}$. For both the almost 
 non-interacting and the strongly attractive case there is an approximate 
 increase by a factor of 2, but in the strongly repulsive case there is 
 a factor of more than 6.     

 Compared to the analysis of the radial pair densities in Sec.~\ref{subsec:raddens} 
 it is evident that the absolute wave-function analysis reveals much more 
 subtle details. In the case of radial pair densities there was the clear 
 trend that improving the level of description leads to an increasing 
 shift of probability from the maxima towards large separations. Similarly, 
 the energies were uniformly lowered. (Remind, however, that the energy
 analysis for $^6$Li$^7$Li showed that such a uniform trend is not found for 
 all heteronuclear systems.) The cuts through the full wave functions show 
 that the effects of coupling and anharmonicity are not as trivial. 
 Most importantly, they indicate that there is a lot of changes of the 
 wave functions for short internuclear separations where, e.\,g., a 
 pseudopotential approach is questionable. The relative importance of this 
 regime of interatomic separations is, however, reduced, if an average over the 
 angles is performed; simply because it scales with the radial part of the 
 volume element, $r^2$. This is also the reason why the energies are not 
 very sensitive to this short-range regime and thus the pseudopotential 
 approach may rather successfully predict also energy differences between 
 different levels of approximation. 

 The wave functions (not their differences) were also used in order to 
 assure that the parameters chosen in this work allow a discussion in terms 
 of a single site of an optical lattice. Different cuts through the 
 wave functions (in different directions relative to the optical lattice) 
 never indicated a substantial wave function amplitude close to the 
 boundaries of the single lattice site.  

 \subsection{Comparison to experiment}
 \label{subsec:compthexp}
\begin{figure}[ht]
 \centering
\includegraphics[width=8.5cm,height=7.0cm]{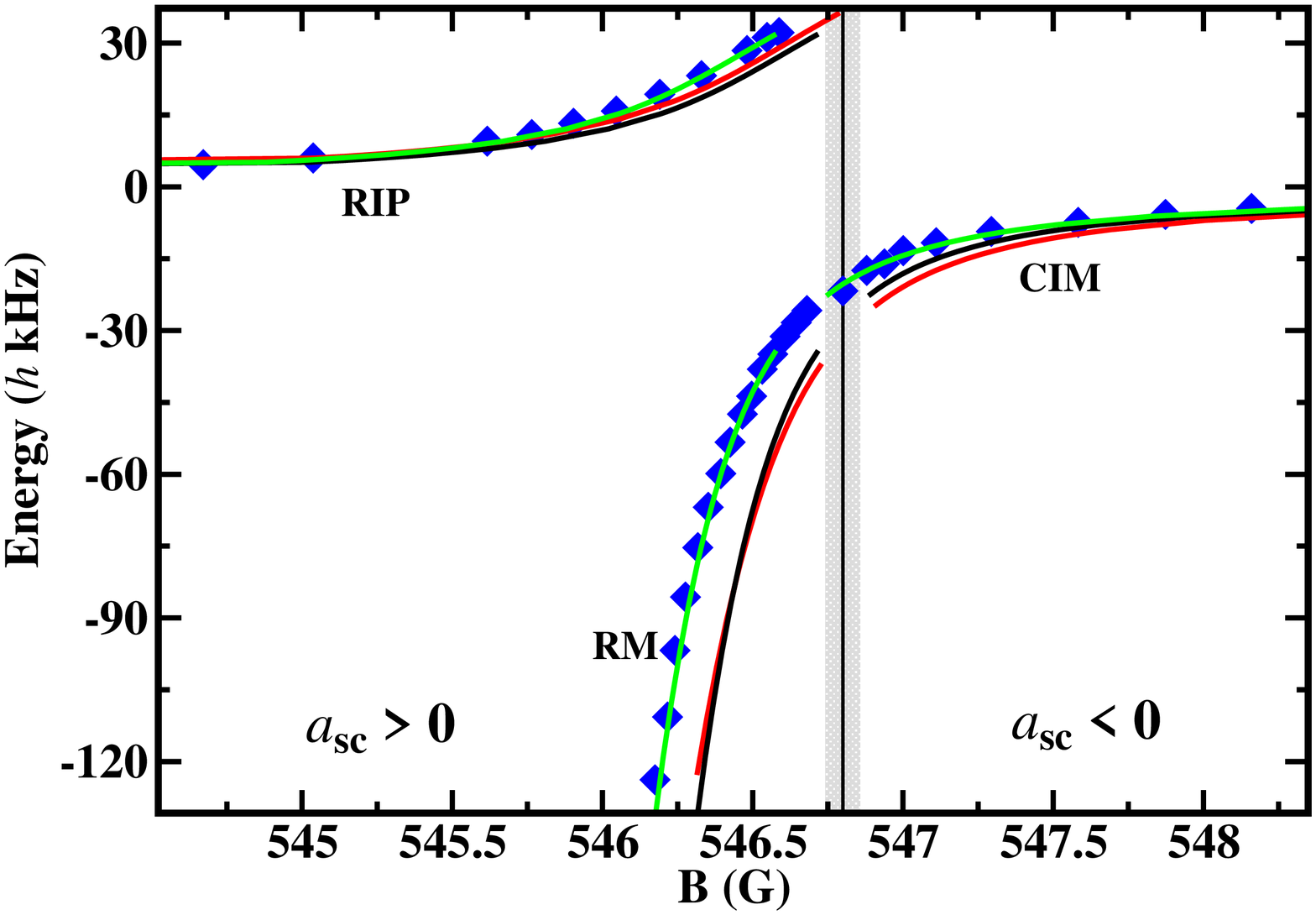}
 \caption{{\footnotesize
     (Color online) 
     The experimentally measured binding energy (diamonds) 
     of \RbK in an optical lattice ($40\,E_r^{\rm{Rb}}$,  
     $\lambda$ of 1030\,nm)  
     together with the theoretically calculated ones for the 
     sextic potential and the energy-independent (black solid) or  
     energy-dependent (red solid) scattering length and the 
     Feshbach resonance parameters $B_0=546.8$ G and $\Delta B = -3$~G.  
     The figure also shows the binding 
     energy for the sextic trap and the energy-independent scattering 
     length for the alternative value $B_0=546.66$\,G (green solid). 
     (All theoretical curves are full CI solutions.)
} }
\label{fig:Bfield}
\end{figure}

 A natural application of the present approach is to model the experimental
 results of C. Ospelkaus et al.~\cite{cold:ospe06a}. In that experiment
 rf association was used to create molecules from fermionic $^{40}$K 
 and bosonic $^{87}$Rb atoms in a 3D cubic optical lattice. The binding
 energy of the heteronuclear molecules was measured as a function of the
 strength of an applied magnetic field. Figure~\ref{fig:Bfield} shows the
 experimental data for a lattice with depth $V_{\rm{Rb}}=40\,E_r^{\rm{Rb}}$ 
 and wavelength $\lambda= 1030\,$nm. 

 Note, the binding energies measured in the
 experiment are not the usual ones. In free space, real molecules~(RM) close
 to the Feshbach resonance exist only on the repulsive side of the resonance
 ($a_{\rm sc}>0$). The binding energy measured in a trap-free situation is
 the one relative to the threshold energy of the continuum. In the
 presence of an external optical lattice this continuum is discretized, and 
 there is instead a first trap-induced state. On the attractive side of the
 resonance ($a_{\rm sc}<0$) the energy of this state is lowered relative to
 the field-free position. This leads to confinement-induced molecules 
 (CIM)~\cite{cold:stoe06}. In the experiment reported in~\cite{cold:ospe06a},
 the binding energy of the RM and CIM in a trap were measured. For 
 $a_{\rm sc}>0$ the excitation energy of the repulsively interacting bound 
 pair~(RIP)~\cite{cold:wink06} where repulsion between bosons and fermions
 shifts the two-particle ground state towards a higher energy was also
 measured. The corresponding~RM,~CIM~and~RIP branches are denoted in
 Fig.~\ref{fig:Bfield} in which the experimental results
 of~\cite{cold:ospe06a} are reproduced for comparison.

\begin{figure}[ht]
 \centering
\includegraphics[width=8.5cm,height=7.0cm]{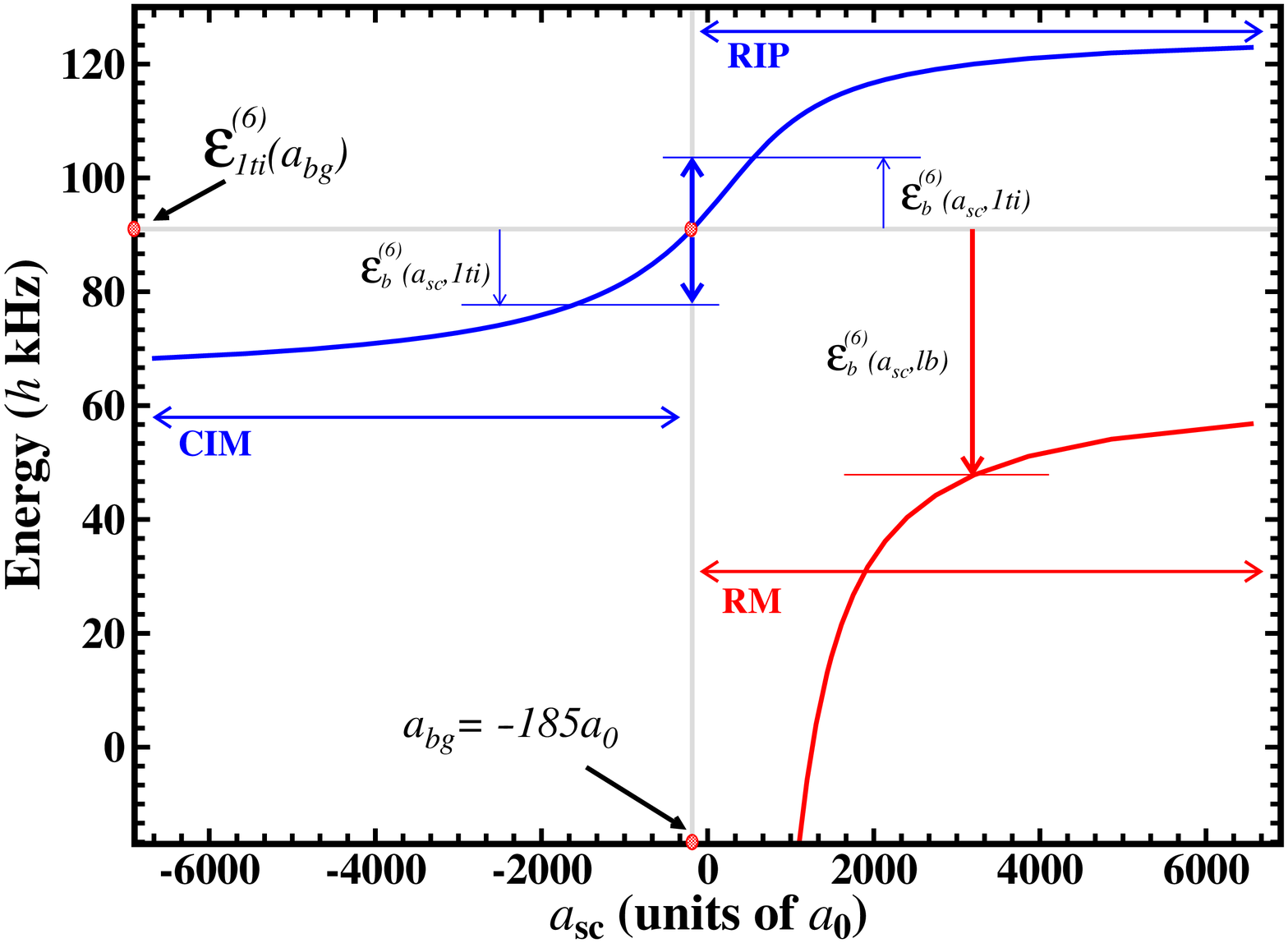}
 \caption{{\footnotesize
     (Color online) 
     Sketch of the procedure for obtaining 
     binding energies from the model. The 1st trap-induced 
     (blue) and least bound (red) state energies obtained with a 
     full CI calculation for the sextic potential (the same as
     in Fig.~\ref{fig:RbK_27_40}(b)) are shown. While the energy-offset 
     of the first trap-induced level (blue arrows) relative to 
     the energy zero ($\eci^{(6)}_{\rm 1ti}(a_{\rm bg})$)  
     is responsible for the confinement phenomena, the 
     energy offset of the least bound state (red arrow) is responsible 
     for the pure molecular ones.
} }
\label{fig:bindingsketch}
\end{figure}

 In order to compare the experimentally measured binding energies with the
 theoretically calculated ones a proper mapping must be
 applied. Figure~\ref{fig:bindingsketch} outlines the procedure of how the
 binding energies were determined in the model. The scattering length 
 $a_{\rm bg}=-185\,a_0$ is chosen as the $B$-field-free background scattering
 length. The energy of the first trap-induced state obtained with the full
 sextic solution at $a_{\rm bg}$ is chosen as energy zero and is marked
 explicitely in Fig.~\ref{fig:bindingsketch}. A variation of the scattering
 length leads to energy shifts of the least bound and the first trap-induced
 states relative to this energy zero. The binding energy is a function of this
 shift as is indicated by arrows in Fig.~\ref{fig:bindingsketch}. Specifically, 
 the binding energy as a function of the scattering length may be obtained from 
 the present theoretical data with help of the relation
 \begin{equation}
   \ds
   \mathbf{E}_{\rm b}^{(n)}(a_{\rm sc};i) =
   \mathbf{E}_{\rm 1ti}^{(n)}(a_{\rm bg})-
   \mathbf{E}_{i}^{(n)}(a_{\rm sc})\quad 
 \end{equation}
 where $i$ and $n$ are, as before, $i=\{{\rm lb,1ti}\}$ and $n=\{{2,6}\}$.
 Furthermore, $\mathbf{E}$ stands for the state energy at a given level of
 approximation $\mathbf{E}=\{E,\eci\}$. This definition of a binding energy
 results in three different branches. While the first trap-induced state is
 responsible for the RIP and the CIM branch, the least bound state is
 responsible for the RM part. The corresponding branches are indicated in
 Fig.~\ref{fig:Bfield} and in the sketch of Fig.~{\ref{fig:bindingsketch}}.

 Experimentally, the binding energies were measured as functions of the magnetic
 field while theoretically calculated energies are functions of the
 interaction strength represented by the scattering length (see
 Sec.~\ref{subsec:manofinin} for details). To provide a $B$ dependence of the
 theoretical data $a_{\rm sc}$ is mapped onto the magnetic field using a
 two-channel approximation~\cite{cold:koeh06} by the aid of
 \begin{equation}
   \label{eq:twochan}
   \ds
   a_{\rm sc}(B)=a_{\rm bg} \left(1-\frac{\Delta B}{B-B_0}\right)
 \end{equation}
 where $\Delta B$ is the resonance width and $B_0$ is the resonance position.
 Equation~(\ref{eq:twochan}) gives in turn for the $B$ field as a function of
 $a_{\rm sc}$
 \begin{equation}
   \label{eq:twochaninv}
   \ds B(a_{\rm sc}) = 
   \Delta B\left(1-\frac{a_{\rm sc}}{a_{\rm bg}}
           \right)^{-1} + B_0\quad .
 \end{equation}
 The $a_{\rm sc}$ values obtained from theory are inserted into 
 Eq.~(\ref{eq:twochaninv}) to determine the $B$ dependence of the energy. 

 Figure \ref{fig:Bfield} shows the binding energy obtained from the full
 sextic solution $\eci^{(6)}$ for the experimental parameters of the trap and
 magnetic field Feshbach resonance parameters 
 $\Delta B =-3$~G~\cite{cold:zacc06} and $B_0=546.8$~G~\cite{cold:ospe06b}. As
 is evident from Fig.~\ref{fig:Bfield} the model does not perfectly agree with
 the experiment. Some possible reasons of the disagreement are discussed in
 the following paragraph.

\subsubsection{Possible reasons of deviation between theory and experiment}
\label{subsec:deviations}
 Equation (\ref{eq:twochan}) is derived for the lattice-free situation under
 the assumption that the collision between two atoms can be approximated by a
 two-channel scattering model. In general, as was already mentioned in
 Sec.~\ref{subsec:manofinin}, the correct theoretical description requires a
 multi-channel scattering treatment which in the present case would  have to
 incorporate also the optical lattice. Moreover, the present model uses
 an "artificial" variation of the scattering length (see
 Sec.~\ref{subsec:manofinin}), and the $a_{\rm sc}$ values obtained from this
 variation are the ones of a single-channel approach.

 Even assuming the validity of Eq.~(\ref{eq:twochan}) there is another
 important factor influencing the comparison of theory and experiment. The
 values of the scattering length $a_{\rm sc}$ and $a_{\rm bg}$ of
 Eq.~(\ref{eq:twochan}) are determined in a lattice-free situation. In the 
 presence of a trap these values must be revised and adjusted to the trap
 parameters. It was shown in~\cite{cold:blum02} that the use of an 
 energy-dependent scattering length $a_{\rm sc}^E$ gives almost correct energy
 levels for two harmonically trapped atoms. The evaluation of $a_{\rm sc}^E$
 requires to solve the complete scattering problem and thus $a_{\rm sc}^E$ can
 only be obtained from the knowledge of the solution for the realistic
 atom-atom interaction potential. Eventually, the trap-free values of the
 scattering length $a_{\rm sc}$ and $a_{\rm bg}$ must be substituted by
 appropriate energy-dependent scattering length $a_{\rm sc}^E$
 values. However, the problem is that the energy-dependent scattering length
 approach is so far developed only for the harmonic approximation, for 
 s-wave collisions, and the uncoupled problem. An anharmonic, for example
 "sextic", energy-dependent scattering length concept as well as any other
 extensions of it do so far not exist to the authors' knowledge.

 In view of the absence of an $a_{\rm sc}^E$ beyond the uncoupled harmonic 
 approximation, 
 the following procedure was adopted. The energy-dependent values
 of the scattering length are obtained using a solution for the
 pseudopotential energy and valid for a harmonic
 trap~\cite{cold:deb03}
\begin{equation}  
\ds \frac{\Gamma\left(\ds -\frac12\frac{\epsilon_{\rm 1ti}^{(2)}}
                {\omega_{\rm ho}}+\frac34\right)}
         {\Gamma\left(\ds -\frac12\frac{\epsilon_{\rm 1ti}^{(2)}}
                {\omega_{\rm ho}}+\frac14\right)}
\ds      =\frac{a_{\rm ho}}{a_{\rm sc}^E\sqrt{2}} \quad ,
\label{eq:energy_pseudo}
\end{equation}
 where $\Gamma$ is a gamma function. The energy of the REL motion obtained for
 the {\it harmonic} trap are imposed into Eq.~(\ref{eq:energy_pseudo}) and the 
 $a_{\rm sc}^E$ values are obtained. The new values of the scattering
 length obtained with this manipulation are used for the mapping of the binding
 energies of $\eci^{(6)}$ with the help of Eq.~(\ref{eq:twochan}). 
 Figure~\ref{fig:Bfield} shows the result of this procedure. As is seen
 from the figure the shift of the spectral curve for the case of the
 energy-dependent scattering length along the $B$-axis is
 not big but the curve is shifted along itself for the RIP branch and is tilted
 for the other ones in immediate proximity to the resonance. This may be seen  
 as an indication that the energy dependence of the scattering length
 (properly included) does not have a too big effect, but the approximate 
 implementation is certainly not conclusive and thus cannot exclude a 
 possible importance.  
\begin{figure}[ht]
 \centering
\includegraphics[width=8.5cm,height=7.0cm]{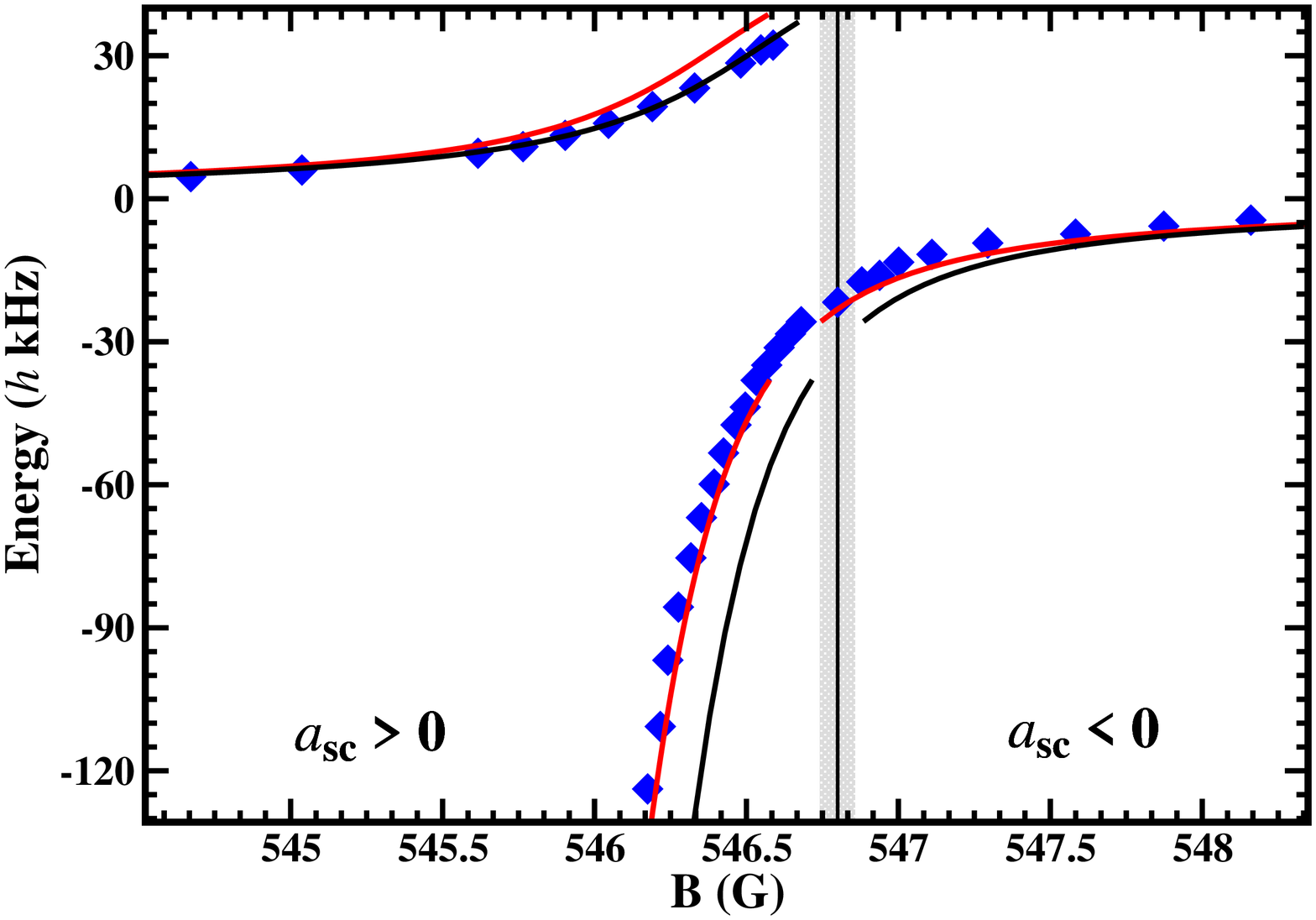}
 \caption{{\footnotesize
     (Color online) 
     As Fig.~\ref{fig:Bfield}, but using a {\it harmonic}
     potential, an energy-independent scattering length,  
     and the Feshbach-resonance positions $B_0=546.8$~G (black solid) 
     or $B_0=546.66$~G (red solid).
} }
\label{fig:BfieldHO}
\end{figure}

 Another important reason of the mismatch between theory and experiment could
 be an insufficient knowledge of the resonance parameters~\cite{cold:deur08}. It
 turns out to be sufficient to change the center of the Feshbach resonance 
 to the value
 $B_0=546.66$~G to match the experimental and the theoretical data. A 
 variation of $B_0$ of this size is well within the experimental 
 uncertainty with which the resonance parameters are known~\cite{cold:ospe06b}. 
 The result obtained with this modified value of $B_0$ is also shown in
 Fig.~\ref{fig:Bfield}. Remarkably, if both parameters $B_0$ and $\Delta B$ 
 are used
 together to fit the experimental curve it leads to a larger error than if only
 the parameter $B_0$ is varied (see the discussion in the following subsection, 
 especially Fig.~\ref{fig:RIPRMCIM}). While a
 variation of $\Delta B$ and $B_0$ shifts the theoretical data along the
 magnetic-field axis, the variation of $a_{\rm bg}$ leads in addition to a
 shift along the energy axis, since it changes the B-field-free energy zero.

 Finally, one may address the question whether despite the number of
 uncertainties the effect of the anharmonicity and coupling is visible in the
 experiment~\cite{cold:ospe06a}. Figure~\ref{fig:BfieldHO} shows the binding
 energies obtained from the harmonic approximation. While the harmonic
 approximation predicts the binding energy of the repulsively-interacting-pair
 part of the spectrum correctly, for other parts it results in a
 disagreement. A variation of the MFR parameters does not lead to a   
 simultaneous matching of all spectral branches. Therefore, it is possible to 
 conclude that in the experiment~\cite{cold:ospe06a} effects of anharmonicity 
 and coupling (and thus deviations from a simple uncoupled harmonic model) 
 were very likely detected. 

\subsubsection{Comparison to a previous theoretical study}
 The effects of anharmonicity and coupling of COM and REL motion in a single 
 site of an optical lattice were also the subject of a recent theoretical 
 study by Deuretzbacher {\it et al.}~\cite{cold:deur08}. 
 The approach therein differs from the present one, since (i) it does not 
 use the full interatomic interaction potential but resorts to the 
 pseudopotential approximation, (ii) a different partitioning of the 
 Hamiltonian is adopted, and (iii) different basis functions
 (eigensolutions of the harmonic oscillator) were adopted. 

 The two independently developed approaches provide the possibility 
 to further check whether theory has achieved a sufficient accuracy 
 to investigate the small deviations from the simple uncoupled harmonic 
 approximation claimed to be found in the experiment in~\cite{cold:ospe06a}.     
 A consequence of difference (i) between the two approaches is furthermore 
 the ability to investigate the adequacy of the pseudopotential adopted 
 in~\cite{cold:deur08}. As a consequence of (iii) the approach 
 in~\cite{cold:deur08} can only be applied to very deep lattices and 
 an extension to multiple-site lattices or even to shallow
 lattices is not straightforward. The reason is the rather strong 
 spatial confinement of the harmonic-oscillator solutions. As a consequence, 
 it needs an impractical large number of basis functions in order to 
 cover an extended spatial regime. Since anharmonicity and coupling 
 effects are different for shallower lattices as is discussed in 
 Sec.~\ref{subsec:othsys}, the tunneling effects may also play an important 
 role~\cite{cold:foel07,cold:trot08}. Within the present approach 
 calculations for multiple-wells and shallow lattices
 are straightforward and were already recently performed~\cite{cold:schn09}.

 The spatial compactness of the harmonic-oscillator eigenfunctions is 
 on the other hand evident from the convergency study with respect to the 
 Taylor expansion of the optical lattice performed in~\cite{cold:deur08}. 
 As was discussed in Sec.~\ref{subsec:spectrum}, such a study is not senseful, 
 since, e.\,g., even-order expansions lead to unphysical continua. Clearly, 
 only a basis that does not explore the corresponding regime of the 
 configuration space does not show any signs of these unphysical 
 continua.

\begin{table}
  \caption{Influence of different levels of approximation
    on the energy of the 1st trap-induced state for 
    three heteronuclear systems. All results are obtained  
    for $a_{\rm sc}=6500\, a_0$ 
    (or $\xi({\rm RbK})=3.34$, 
        $\xi({\rm LiCs})=3.76$,
        $\xi({\rm LiLi})=3.24$),  
    lattice depths of
    $V_1=V_2=10\,E_{\rm{r,rel}}$ where $E_{\rm r,rel}=k^2/(2\mu)$, 
    and a wavelength $\lambda=1000$\,nm. }
  \begin{ruledtabular}
     \begin{tabular}{lccccc}
        atom pair & $E_{\rm 1ti}^{(2)}$ & $\Delta_2$ &
        $\eci^{(6)}_{\rm 1ti}-\eci^{(2)}_{\rm 1ti}$  & $\Delta_{\rm tot}$ \\
         \vspace{-0.3cm}
         \;& \;& \;& \;& \; \\
        \hline
         \vspace{-0.3cm}
         \;& \;& \;& \;& \; \\
         \RbK  [present]\;           & 3.79 &  -0.12 &  -0.29  & -0.41  \\
         \phantom{\RbK}  \cite{cold:deur08}\;  & 3.74 &  -0.12 &  -0.27  & -0.39  \\
         \LiCs [present] \;          & 2.93 &  -0.38 &  -0.22  & -0.60  \\
         \phantom{\LiCs} \cite{cold:deur08}\;  & 2.88 &  -0.35 &  -0.22  & -0.57  \\
         \LiLi [present]\;           & 3.93 &  -0.01 &  -0.30  & -0.31  \\
         \phantom{\LiLi} \cite{cold:deur08} \; & 3.92 &  -0.01 &  -0.29  & -0.30  \\
     \end{tabular}
  \end{ruledtabular}
  \label{tab:pfancomp}
\end{table}
 Table~\ref{tab:pfancomp} shows a comparison of some energies and 
 energy differences obtained with the numerical approach in~\cite{cold:deur08} 
 and the present one for a large positive scattering length 
 ($a_{\rm sc}=6500\,a_0$). The results obtained with the two approaches 
 do not differ very much in the case of all three 
 considered alkali-metal dimers. The agreement of the energy differences 
 is overall slightly better than the one of the absolute energies. The 
 comparison seems to confirm the proper numerical implementation of both 
 numerical approaches. Most importantly, it demonstrates that for the 
 calculation of energy shifts as well as anharmonic and coupling effects 
 in a single site of an optical lattice the pseudopotential approach  
 remains valid; at least to a very good approximation. 

 It is presently not possible to  
 attribute the remaining differences to the different atomic interaction 
 potential or some remaining numerical uncertainty. Note, the different 
 interaction potential influences the results in two ways. First, the 
 $\delta$-type pseudopotential does not properly account for the short-range 
 part of the interaction. Second, the mapping of the energy to a corresponding 
 interaction strength is different in the two approaches. In the 
 pseudopotential approach the scattering length is simply a parameter that 
 enters the interaction potential, while it is extracted from the resulting 
 wave function as was described in Sec.~\ref{subsec:manofinin}. As a 
 consequence, there is a finite range in which $a_{\rm sc}$ can be varied 
 within the present approach.  

\begin{figure}[ht]
 \centering
\includegraphics[width=8cm,height=18cm]{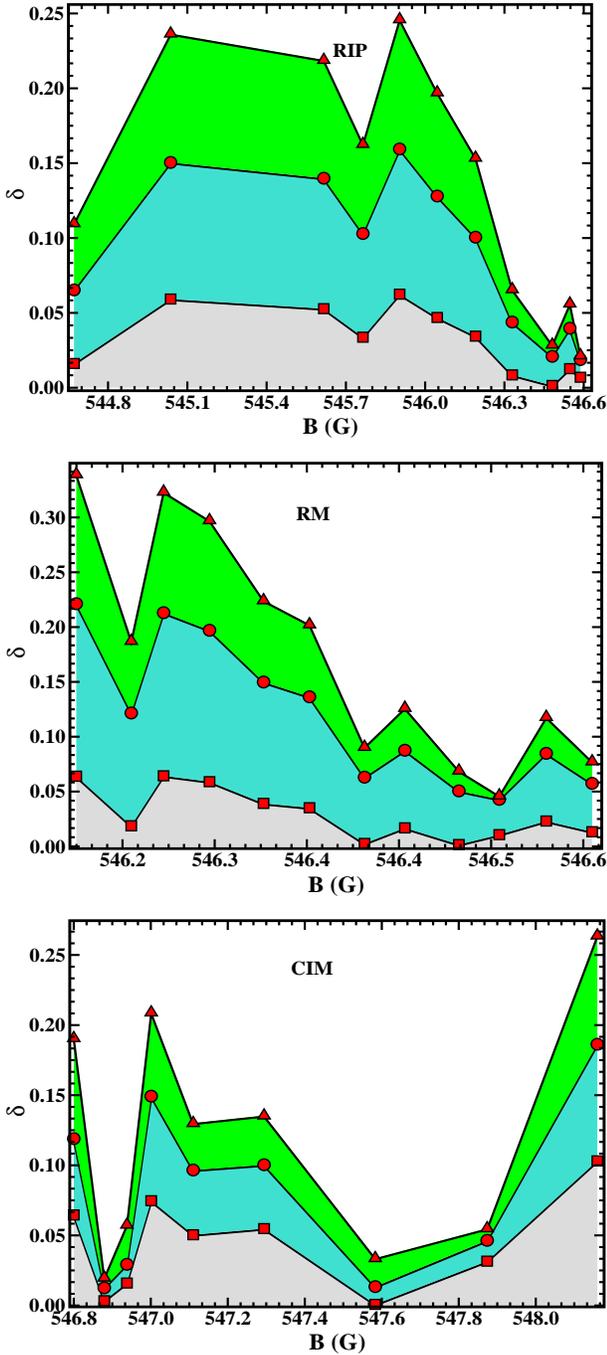}
 \caption{{\footnotesize
     (Color online)
     The relative error defined in Eq.~(\ref{eq:deltatheoexp})
     for the alternative Feshbach parameters
     $B_0=546.660$~G, $\Delta B = -3$~G (squares),
     $B_0=546.669$~G, $\Delta B = -2.92$~G (circles), and
     $B_0=546.66$~G, $\Delta B = -2.92$~G (triangles).
} }
\label{fig:RIPRMCIM}
\end{figure}

 Comparable to the present finding (see Sec.~\ref{subsec:deviations}) the 
 binding-energy spectrum of \RbK calculated in~\cite{cold:deur08} does 
 not agree very well with the experimental one in~\cite{cold:ospe06a}, 
 if the previously experimentally determined Feshbach-resonance parameters 
 ($B_0=546.8$ G, $\Delta B = -3$~G) are used. The authors 
 in~\cite{cold:deur08} proposed that with the aid of 
 the calculation it is in fact possible to improve on the MFR parameters. 
 Such a fit (with the energy-independent scattering length) yielded the 
 new resonance parameters $B_0=546.669$~G and 
 $\Delta B = -2.92$~G~\cite{cold:deur08}. This has to be contrasted 
 with the present fit that yields the new resonance position 
 $B_0=546.660$~G but an unchanged width ($\Delta B = -3$~G), as was 
 discussed in Sec.~\ref{subsec:manofinin}. Thus there is a similar (though 
 slightly larger) trend for $B_0$, but disagreement with the results 
 in~\cite{cold:deur08} with respect to $\Delta B$.  

 In view of the different fit results, it is important to investigate in 
 more detail their sensitivity on the fit parameters.  
 The quality of the fit depends on the agreement between the calculated 
 binding energy ($\eci^{(6)}(B)$) and the experimental one ($E_{\rm exp}$). 
 It is thus given by the relative error
\begin{equation}
  \ds 
  \delta(B) = \left|
  \frac{E_{\rm exp}(B)-\eci^{(6)}(B)}{E_{\rm exp}(B)} \right| \quad .
  \label{eq:deltatheoexp}
\end{equation}
 Figure~\ref{fig:RIPRMCIM} shows $\delta (B)$ for three sets of MFR 
 parameters: (i) $B_0=546.660$~G and $\Delta B = -3$~G (optimal 
 fit parameters, this work), (ii) $B_0=546.669$~G and $\Delta B = -2.92$~G 
 (optimal fit parameters found in~\cite{cold:deur08}), and 
 (iii) $B_0=546.660$~G and $\Delta B = -2.92$~G 
 (optimal fit parameter found in this work for $B_0$, but 
 $\Delta B$ from~\cite{cold:deur08}).  
 As is evident from Fig.~\ref{fig:RIPRMCIM}, any variation of either 
 $\Delta B$ or $B_0$ from their optimal values results in an increased error 
 for all energy branches and all magnetic fields. Clearly, the fit shows a 
 well defined minimum and thus there is no ambiguity in the fit parameters 
 as it could occur, e.\,g., in the case of very shallow minima where 
 the outcome of the fit may be determined by small numerical inaccuracies.  

 Provided the fit fidelity in~\cite{cold:deur08} is comparable to the 
 present one, i.\,e., a fit with the binding energies calculated 
 in~\cite{cold:deur08} using the optimal fit parameters of the present 
 work would disagree with the experiment in a similarly pronounced fashion 
 as shown in Fig.~\ref{fig:RIPRMCIM}, it is presently impossible to 
 conclude whether theory has already reached the level of accuracy 
 that is required for an improved determination of MFR parameters. 
 While both fits appear to indicate a smaller value of $B_0$ compared 
 to the one previously extracted from experiment, the deviation between 
 both fits is only about half as small as the improvement claimed 
 in~\cite{cold:deur08}. Clearly, such a result is from a statistical 
 point of view inconclusive. In the case of the width $\Delta B$ the present 
 finding agrees even fully to the previously determined value and thus 
 disagrees with the result of the fit in~\cite{cold:deur08}.  

 In order to obtain a more conclusive result it is vital to investigate 
 whether the differences between the results in~\cite{cold:deur08} and 
 the present ones are solely due to the use of the pseudopotential
 approximation or the more realistic interatomic interaction potential 
 in the two studies. If this were the case, the fit results of the 
 present study should be regarded as the more accurate ones. Furthermore, 
 this would be an important example for the need to consider the 
 interatomic interaction on a more accurate level than the one provided 
 by the pseudopotential approximation. Since the implementation of the 
 pseudopotential is due to the singular behavior of the $\delta$ function 
 non-trivial in the context of the present approach, such an investigation 
 has to be postponed to a separate work. Clearly, more experimental data 
 (for different heteronuclear systems) would also be very important for 
 gaining a deeper insight and it is hoped that the present work 
 stimulates such experimental activities.

 Finally, there are two further uncertainties in the determination of the MFR 
 parameters from a fit like the one in~\cite{cold:deur08} or in the present 
 work. They are related to the way in which the mapping of the theoretical 
 data onto the magnetic field is performed. As already mentioned, this 
 mapping is usually based on the assumption of validity of 
 Eq.~(\ref{eq:twochaninv}) and thus on the assumption that the $B$-field 
 mapping of the multichannel MFR can be performed based solely on a 
 scattering-length variation. Even in this case there is, however, 
 the problem of the proper determination of the energy-dependent 
 scattering length in an optical lattice which is so far unknown. 
 The use of $a_{\rm sc}^E$ extracted from the harmonic uncoupled energies 
 for the mapping of the the full sextic energy results effectively in a 
 shift of the energy-independent curve, as is seen in Fig.~\ref{fig:Bfield}. 
 However, both the energy-dependent and energy-independent $a_{\rm sc}$ 
 discussed in this work utilize the same harmonic energy curve  
 ignoring also the coupling to the COM motion. How the situation would 
 change, if $a_{\rm sc}^E$ for a non-harmonic solution would be used, is 
 difficult to predict, since the other curves in Fig.~\ref{fig:RbK_27_40} 
 not only differ in shape, but are also shifted relative to each other and 
 contain the COM part. The overall good agreement of the theoretical 
 binding energies (with fitted MFR parameters) to the experimental data 
 is of course very suggestive that these uncertainties have a small 
 influence, but this may be a pure coincidence. 

 Both, the investigation of the appropriateness 
 of the $B$-field mapping as well as the question of the possibility 
 to define an energy-dependent scattering length beyond the uncoupled 
 harmonic approximation require a theoretical approach for the treatment 
 of two atoms in an optical lattice as the one presented in this work 
 and is presently pursued. It should, however, be emphasized that these 
 uncertainties only affect the analysis of dimers close to a (magnetic) 
 Feshbach resonance or, in general, if the proper description of the 
 atom-pair requires a multi-channel treatment. The results of the 
 previous sections are valid independently of these uncertainties. 
 Different interaction regimes are experimentally accessible within 
 the validity regime of a single-potential-curve treatment even 
 for the same dimer by considering different isotopes or electronic 
 states. The simplicity of experimental tunability as is 
 found for magnetic Feshbach resonances is then of course lost.

 \section{Conclusion and outlook}
 \label{sec:outlook}
 An approach which allows for a full numerical description of two ultracold
 atoms in 3D optical lattice is developed. A detailed analysis of
 anharmonicity and coupling of center-of-mass and relative coordinates  
 in terms of energy values and wave functions was performed for heteronuclear 
 dimers in a single site of an optical lattice. It is explained, why such 
 a single site is optimally described by a sextic potential, if a finite 
 Taylor expansion is used. The effects 
 of deviations from the harmonic approximation and of the coupling were
 quantified and analyzed for different heteronuclear systems, confinement 
 strengths and interatomic interaction regimes. The influence of the 
 lattice is found to be always much stronger for the first trap-induced 
 state than for the least bound state. As a consequence, binding energies 
 are modified by the lattice mainly by the modification of the first 
 trap-induced state.  

 While the energy deviations from the harmonic uncoupled approximation is 
 for all three considered generic heteronuclear dimers largest for strong 
 repulsive interaction, the relative size (and even sign) of the energy 
 change due to coupling or anharmonicity varies for the different dimers. 
 The same is true for the the influence of the trap depth. A deeper lattice 
 can lead to smaller or larger energy differences between the harmonic 
 uncoupled or the full coupled solution. While the analysis of the radial 
 pair densities shows that the lattice mainly influences the maxima located 
 at large interatomic separations, the analysis of cuts through the 
 wave functions in absolute coordinates reveals non-negligible changes 
 also at short interatomic distances. This may have important consequences 
 for the validity of pseudopotential approximations. 

 The results of the present theoretical approach are also compared to 
 a recent experiment in which the binding energies of \RbK have been 
 measured as a function of an external magnetic field tuned close to 
 a magnetic Feshbach resonance. The assumptions necessary for such a 
 comparison are carefully discussed. It is found that very good agreement 
 between experiment and theory can only be reached, if the previously 
 experimentally determined resonance parameters are modified. Since this 
 needed modification is within the error bars with which the parameters 
 had been determined before, this is not only reasonable, but may even 
 indicate the possibility to more accurately determine the width and 
 position of magnetic Feshbach resonances in ultracold atomic gases, 
 as was proposed recently in a comparable theoretical study. However, 
 the resonance parameters determined in the previous study based on 
 the pseudopotential approximation differ from the ones found in the 
 present work. If this deviation is due to the pseudopotential 
 approximation is difficult to judge at this moment. If this were the 
 case, the then found breakdown of the pseudopotential approximation 
 would, of course, be a very interesting finding. A further investigation 
 is therefore of great interest, and the present work stimulates hopefully 
 also further experimental work in this direction. Since the influence of 
 anharmonicity and coupling becomes more important for less deep optical 
 lattices and for excited trap levels corresponding  
 experiments like the ones in \cite{cold:muel07,cold:foel07,cold:fall07} 
 are expected to provide further tests of the approach presented in this 
 work.     

 Since the present approach was rather generally formulated and implemented,  
 it allows immediately for further investigations that have partly 
 been started or are planned for the future. This includes the consideration 
 of highly anisotropic, asymmetric (disordered), or multiple-well lattice 
 geometries. First results for triple-well potentials have, e.\,g., 
 recently been used for the determination of Bose-Hubbard parameters 
 and an investigation of the validity of the Bose-Hubbard model 
 itself~\cite{cold:schn09}. The study of an interesting physical phenomena
 like, e.\,g., the trap-induced resonances~\cite{cold:olsh98,cold:stoc03} are
 also planned. Further extensions of the approach should also allow to study
 the case of a pair of atoms or molecules interacting by non-centric, e.\,g.,
 dipolar interactions. Finally, it is planned to extend the method for studies
 of the time-dependent dynamics of atomic pairs in time-varying lattices.

\section*{Acknowledgments}
 The authors are grateful to K.~Sengstock, K.~Bongs and F.~Deuretzbacher for
 valuable discussion and to the {\it Stifterverband f\"ur die Deutsche
 Wissenschaft}, to the {\it Deutsche Forschungsgemeinschaft} (within 
 {\it Sonderforschungsbereich} SFB 450), and to 
 the {\it Fonds der Chemischen Industrie} for financial support.

% \bibliography{cold,gen,vdw,bsp}

%
 \end{document}